\documentclass[%
 aip,
 amsmath,amssymb,
preprint,%
]{revtex4-1}

\usepackage{graphicx}
\usepackage{dcolumn}
\usepackage{bm}

\usepackage[utf8]{inputenc}
\usepackage[T1]{fontenc}
\usepackage{mathptmx}
\usepackage{epstopdf, epsfig}
\usepackage{natbib}
\usepackage{xcolor}
\usepackage{array, booktabs, makecell}
\usepackage{siunitx, mhchem}
\usepackage{amsfonts}
\usepackage{amsmath}
\usepackage{amssymb}
\usepackage{hyperref}
\graphicspath{{./figure/}}
\newcommand{\bfv}[1]{{\bf #1}}
\newcommand{\dd}{\text{\,d}}
\newcommand{\bfvg}{\bfv{g}}
\newcommand{\bfvW}{\bfv{W}}
\newcommand{\bfvK}{\bfv{K}}
\newcommand{\bfvC}{\bfv{C}}

\newcommand{\bfvPhi}{\bm {\Phi}}
\renewcommand{\l}{\left}
\renewcommand{\r}{\right}
\newcommand{\bfvM}{\bfv{M}}
\newcommand{\bfvF}{\bfv{F}}
\newcommand{\bfvx}{\bfv{x}}
\newcommand{\bfvu}{\bfv{u}}
\newcommand{\bfvv}{\bfv{v}}
\newcommand{\bfvO}{\boldsymbol{\Omega}}

\begin{document}

\title{Data Driven Learning of Mori-Zwanzig Operators for Isotropic Turbulence}

\author{Yifeng Tian}
\email{yifengtian@lanl.gov}
 \affiliation{Computational Physics and Methods Group, Computer, Computational and Statistical Sciences Division (CCS-2), Los Alamos National Laboratory, Los Alamos, NM 87545, USA}

\author{Yen Ting Lin}
\affiliation{Information Sciences Group, Computer, Computational and Statistical Sciences Division (CCS-3), Los Alamos National Laboratory, Los Alamos, NM 87545, USA
}%
\author{Marian Anghel}
\affiliation{Information Sciences Group, Computer, Computational and Statistical Sciences Division (CCS-3), Los Alamos National Laboratory, Los Alamos, NM 87545, USA
}%
\author{Daniel Livescu}%
\affiliation{Computational Physics and Methods Group, Computer, Computational and Statistical Sciences Division (CCS-2), Los Alamos National Laboratory, Los Alamos, NM 87545, USA}%

\date{\today}

\begin{abstract}
Developing reduced-order models for turbulent flows, which contain dynamics over a wide range of scales, is an extremely challenging problem. In statistical mechanics, the Mori-Zwanzig (MZ) formalism provides a mathematically formal procedure for constructing reduced-order representations of high-dimensional dynamical systems, where the effect due to the unresolved dynamics are captured in the memory kernel and orthogonal dynamics. Turbulence models based on MZ formalism have been scarce due to the limited knowledge of the MZ operators, which originates from the difficulty in deriving MZ kernels for complex nonlinear dynamical systems. In this work, we apply a recently developed data-driven learning algorithm, which is based on Koopman's description of dynamical systems and Mori's linear projection operator, on a set of fully-resolved isotropic turbulence datasets to extract the Mori-Zwanzig operators. With data augmentation using known turbulence symmetries, the extracted Markov term, memory kernel, and orthogonal dynamics are statistically converged and the Generalized Fluctuation-Dissipation Relation can be verified. The properties of the memory kernel and orthogonal dynamics, and their dependence on the choices of observables are investigated to address the modeling assumptions that are commonly used in MZ-based models. A series of numerical experiments are then constructed using the extracted kernels to evaluate the memory effects on predictions. Results show that the prediction errors are strongly affected by the choice of observables and can be further reduced by including the past history of the observables in the memory kernel.
\end{abstract}

\maketitle

\section{\label{sec:intro}Introduction}
Direct and accurate computation of complex nonlinear dynamical systems in physical sciences and engineering applications, such as turbulent flows, are in general prohibitively expensive due to the existence of wide range of length and time scales. This challenge motivated the development of reduced order models (ROM) to achieve fast and efficient solutions in practical applications. In the field of turbulence simulations, Reynolds-Averaged Navier-Stokes (RANS) and Large Eddy Simulations (LES) have been widely adopted as alternatives for Direct Numerical Simulation (DNS). The computational complexity is reduced through coarse-graining, which effectively narrows the ranges of scales that needs to be resolved \citep{sagaut2006large}. High-order moments and their dynamical equations, which account for the effects from unresolved scales, naturally emerge during the coarse-graining process. The high-order moments are then truncated and surrogate models, commonly referred to as subgrid-scale models, have been developed to account for the effects from the high-order moments contributions. Such surrogate models are usually developed under the assumptions of scale similarity and homogeneity, and the resulting simplified models are usually Markovian in nature. \citet{kraichnan1964decay, kraichnan1965lagrangian}  introduced Direct Interaction Approximation (DIA) as a non-Markovian closure model for turbulence statistics. The evolution of a new quantity, the infinitesimal response function, is employed to model the response of turbulent flows to infinitesimal perturbations. However, this method suffers from certain theoretical (inability of reproducing inertial range behavior) and practical (difficulty in calculating long-time statistics) weaknesses \citep{bos2013lagrangian}.

The Mori-Zwanzig (MZ) formalism was originally developed in statistical physics for constructing low-dimensional, non-Markovian models for high-dimensional nonlinear dynamical systems \citep{mori1965transport,zwanzig1973nonlinear}. It provides a mathematically exact procedure for developing reduced-order models for high-dimensional systems. The resulting lower-dimensional model, referred to as the Generalized Langevin equation (GLE), consists of a Markovian term, a memory term, and a noise term. The GLE as derived in the framework of MZ formalism is an exact representation of the dynamics of the reduced-order model. In the context of RANS and LES modeling, the truncated high-order moments in the high-dimensional closed dynamical systems can be accounted for in the memory integral and noise terms under the Mori-Zwanzig formalism. However, in most previous efforts to subgrid-scale modeling, the evolutionary equations are usually treated as Markovian. 

Modeling turbulence under the MZ formalism is extremely challenging due to the limited understanding of the memory kernels and orthogonal dynamics. Obtaining the structure of the memory kernel requires the solution of the unresolved orthogonal dynamics, which is another high-dimensional nonlinear dynamical system. Theoretically obtaining the memory kernel is a very difficult task especially for complex nonlinear system. Despite this difficulty, the optimal prediction framework developed in \citet{chorin2000optimal,givon2005existence,chorin2007problem} provided a formal procedure for analyzing the memory effects and developing surrogate models based on the MZ  formalism. One of the major modeling difficulties is determining the structure and the length of the memory integral. The widely used t-model approximates the memory length as equal to simulation time $t$ and has been used by \citet{bernstein2007optimal,hald2007optimal,chandy2010t} for prediction of Burger's equations, Euler equations, and Navier-Stokes (N-S) equations. \citet{parish2017dynamic} later proposed a dynamic-$\tau$ model to approximate the memory length using the similarity between two coarse-graining levels. The renormalized MZ models \citep{stinis2013renormalized,stinis2015renormalized} embedded a larger class of models that share similar functional forms with MZ formalism but with different coefficients to approximate the memory integral. \citet{stinis2007higher,parish2017non} further used finite order expansion of the orthogonal dynamics and cast it to a set of differential equations that represent the effects of memory integral with finite memory length. These MZ-based turbulence models rely on simplified assumptions and observations of the turbulent flow, most of which have not been verified due to the difficulty in deriving or extracting the MZ operators. \citet{gouasmi2017priori} proposed a method for estimation of the memory integrals using pseudo orthogonal dynamics, which is only exact for linear dynamics.

 The above mentioned models approach the dimensional reduction starting with the original nonlinear systems. The challenges in approximating the memory kernel come from the non-linearity of the equations. For the same dynamical system, there exists another formulation as proposed by \citet{koopman1931hamiltonian,koopman1932dynamical}. In Koopman's description, the system is characterized by a collection of observables which are functions of the original phase space coordinates. The Koopmanian formulation describe how observables evolve in an infinite-dimensional Hilbert space, which is composed of all the possible observables. The advantage of this formulation is that the evolution of the observables, which is a vector in the infinite dimensional Hilbert space, is always linear, even for systems that are nonlinear in the phase-space picture. The disadvantage of this formulation is that the state space of the system, which consists of
all possible observables, is infinite dimensional. Based on the Koopman description of dynamical system, approximate learning methods, such as Dynamic Mode Decomposition (DMD) \citep{schmid2010dynamic} and Extended Dynamic Mode Decomposition (EDMD) \citep{williams2015data} have been developed for data-driven modeling of dynamical systems. By combining the Koopman description with MZ formalism, it is possible to perform a dimensional reduction of the infinite dimensional Koopmanian linear formulation to a finite, low-dimensional dynamical system with memory kernels and orthogonal dynamics. Since the observables evolve in a linear space, the learning problem is convex, which can greatly simplify the learning of MZ operators. \citet{lin2021data} proposed a data driven learning framework for MZ memory kernel and noise term under the generalized Koopman formulation, and analyzed the properties of these terms for a Lorenz `96 model. This is the first study that successfully extracted the MZ operators for a nonlinear dynamical system.

In this work, we take the first step to apply the learning algorithm in \citet{lin2021data} to a homogeneous isotropic turbulence DNS database to extract the Markov, memory, and orthogonal (noise) terms for the coarse-grained Navier-Stokes system. To the authors' best knowledge, there has been no study using data-driven methods to accurately extract MZ terms for Navier-Stokes turbulence, despite the fact that understanding the properties of the memory kernels and orthogonal dynamics are crucially important not only to quantitatively address the assumptions in MZ-based turbulence models but, more generally, understand the past memory effects for NS truncated dynamics. The manuscript is organized as follows: in Section \ref{sec:MZ}, the MZ formalism, as a generalized Koopman learning framework, will be introduced. The data-driven learning framework for MZ kernels is introduced in Section \ref{sec:learning}. The DNS database and post-processing procedures are then explained in Section \ref{sec:dnsdata}. The results and discussions are presented in Section \ref{sec:results} and the conclusions are drawn in Section \ref{sec:conclusion}.

\section{Mori-Zwanzig Formalism}
\label{sec:MZ}
Consider the following semi-discrete high-dimensional Ordinary Differential Equations (ODE) for the full set of state variables $\bfvPhi(t) =\l [ \phi_1(t),...,\phi_N(t) \r ]^T\in \mathbb{R}^N$:
 \begin{equation}
 \label{eqn:ode}
     \frac{d\bfvPhi(t)}{dt}=R(\bfvPhi(t)), \bfvPhi(0)=\bfvx
 \end{equation}
 \noindent where $R: \mathbb{R}^N \rightarrow \mathbb{R}^N$ is a $N$-dimensional vector of nonlinear real functions defined on $\mathbb{R}^N$. Due to the difficulties in simulating and analyzing high-dimensional nonlinear dynamical systems, it is generally desirable to develop low-dimensional representations of the same system. In order to achieve this reduction of complexity, we consider the evolution of a set of observables $\bfvu(\bfvx, t):=\bfvg(\bfvPhi(\bfvx, t))$, where $\bfvg : \mathbb{R}^N \rightarrow \mathbb{R}^D$ is a $D$-dimensional vector of observables (functions of the phase space variables $\bfvPhi(\bfvx, t)$) and, in general, $D<N$. Here, we use $\bfvPhi(\bfvx, t)$ to denote the solution to equation (\ref{eqn:ode}) with the initial conditions $\bfvPhi(0) = \bfvx$, and $\bfvg(\bfvx,t)$ to denote the observables $\bfvg(\bfvPhi(\bfvx, t))$ at time $t$.  One way to define the observables is to decompose the state variables $\bfvPhi$ into resolved/relevant variables $\hat{\bfvPhi}(\bfvx, t) = \l [ \phi_1(t),...,\phi_D(t) \r ]^T\in \mathbb{R}^D$, and unresolved ones $\Tilde{\bfvPhi} = \l [ \phi_{D+1}(t),...,\phi_N(t) \r ]^T\in \mathbb{R}^{N-D}$, and the evolutionary equation for the resolved variables $\hat{\bfvPhi}$ is developed to reduce the dimension of the original nonlinear ODEs. In Sec.~\ref{sec:NS-ROM} we will describe a set of coarse-grained observables which are derived by applying a spatial filter to the velocity field of Navier-Stokes equations.
 For a system with sufficient separation of scales, it might be possible to decompose the system into slow and fast dynamics; the latter might be modeled as function of the former and some simple (white) noise. However, in most physical systems, there exists a continuous spectrum of scales so that the dynamics of the resolved variables are nonlinearly coupled to the unresolved ones. To formally solve this problem, \citet{mori1965transport, zwanzig1973nonlinear} developed the projection-based method to express the effects of the unresolved variables in terms of the resolved ones. The key result of Mori-Zwanzig's formulation of the reduced-dimensional system is the Generalized Langevin Equation (GLE), which is characterized by the emergence of the memory kernel, which is represented as the convolutional integral of the past history of the resolved variables, and orthogonal dynamics, which describes evolution of unresolved variables in the orthogonal space.

\subsection{Derivation of the Generalized Langevin Equation: Key Construct of Mori-Zwanzig Formalism}
\label{sec:GLE}
In this section, we provide a formal derivation of the Mori-Zwanzig formalism. Consider the semi-discrete nonlinear ODE system as shown in Eq.~\eqref{eqn:ode} with the initial conditions $\bfvPhi(0)=\bfvx, \bfvx \in \mathbb{R}^N$. The evolution of a set of observables $\bfvu(\bfvx,t):=\bfvg(\bfvPhi(\bfvx, t))$, where $\bfvg: \mathbb{R}^N \rightarrow \mathbb{R}^D$ is a $D$-dimensional vector of observables, can be posed as a linear Partial Differential Equation (PDE) in the Liouville form:
\begin{equation}
    \label{eqn:pde}
    \frac{\partial}{\partial t} \bfvu(\bfvx,t) = \mathcal{L}\bfvu(\bfvx,t), \quad \bfvu(\bfvx,0) = \bfvg(\bfvPhi(\bfvx, 0))=\bfvg(\bfvx),
\end{equation}
\noindent where $\mathcal{L}$ is the Liouville operator,
\begin{equation}
\mathcal{L}:= \sum_{i=1}^{N} R_i(\bfvx)\partial_{x_i}. 
\end{equation}
\noindent Thus, the PDE (\ref{eqn:pde}) becomes the ODE (\ref{eqn:ode}) along the characteristics curves. A special choice of $\bfvg$ in (\ref{eqn:pde}) is $g_i(\bfvx):=x_i$, $i \in 1,...,D$, that extracts the $i^{th}$ component of the state of the system, $\bfvx$. Using the semigroup notation, the solution of the linear PDE can be written as $\bfvu(\bfvx,t)= e^{t\mathcal{L}}  \bfvg (\bfvx) $, where $e^{t\mathcal{L}}$ is referred to as the evolution operator.  It can be shown that 
$ e^{t\mathcal{L}} \bfvg (\bfvx)  = \bfvg( e^{t\mathcal{L}} \bfvx ) = \bfvg(\bfvPhi(\bfvx, t))$ since $ e^{t\mathcal{L}} \bfvx  = \bfvPhi(\bfvx, t)$. Alternatively, this can be written as $ e^{t\mathcal{L}} \bfvPhi(\bfvx, 0)  = \bfvPhi(\bfvx, t)$ 
and we recognize that the operator $\mathcal{K}_t = e^{t\mathcal{L}}$ is the one parameter family of Koopman operators. We further remark that the evolution operator (Koopman operator) $e^{t\mathcal{L}}$ and the Liouville operator $\mathcal{L}$ commute, that is $\mathcal{L} e^{t\mathcal{L}} 
=  e^{t\mathcal{L}}\mathcal{L}$. Hence, Eq.~\eqref{eqn:pde} becomes
\begin{equation}
    \label{eqn:semigroup}
\frac{\partial}{\partial t} e^{t\mathcal{L}}  \bfvg (\bfvx)  = \mathcal{L} e^{t\mathcal{L}} \bfvg (\bfvx) 
=  e^{t\mathcal{L}}\mathcal{L}  \bfvg (\bfvx) \, .
\end{equation}

In order to construct reduced-order representation of the linear PDE using the reduced $D$-dimensional vector of observables $\bfvg$, with the initial condition $\bfvu(\bfvx, 0)=\bfvg(\bfvx)$, a projection operator, $P$, needs to be specified that maps functions $f(\bfvx)$ defined in the Hilbert space into the subspace $Span\{g_1(\bfvx), ..., g_D(\bfvx)\}$. The particular choice of projection operator determines the functional form of the Mori-Zwanzig formulation. Examples of the projection operators include nonlinear projection operator that relies on the marginalization of the under-resolved observables \citep{zwanzig1973nonlinear} and finite rank projection operator that relies on the inner product in the Hilbert space \citep{mori1965transport}. After the projection operator $P$ is defined, its complement $Q$ is denoted as $Q = I-P$ and satisfies $PQ=QP=0$, where $I$ is the identity operator. We then substitute the Dyson identity \citep{j2007statistical}:

\begin{equation}
    \label{eqn:dyson}
    e^{t\mathcal{L}} = e^{t(P+Q)\mathcal{L}} = e^{tQ\mathcal{L}} + \int_0^t e^{(t-s)\mathcal{L}}P\mathcal{L}e^{sQ\mathcal{L}}ds,
\end{equation}

\noindent in Eq.~\eqref{eqn:semigroup} and arrive at:
\begin{eqnarray}
\frac{\partial}{\partial t} \l [ e^{t\mathcal{L}}\bfvg (\bfvx) \r ] =  e^{t\mathcal{L}} \mathcal{L} \bfvg (\bfvx) =   e^{t\mathcal{L}} P\mathcal{L} \bfvg (\bfvx) +  e^{t\mathcal{L}} Q\mathcal{L} \bfvg (\bfvx) \nonumber \\
= e^{t\mathcal{L}} P\mathcal{L} \bfvg (\bfvx) +   e^{tQ\mathcal{L}} Q\mathcal{L} \bfvg (\bfvx) + \int_0^t  e^{(t-s)\mathcal{L}}P\mathcal{L}e^{sQ\mathcal{L} }Q\mathcal{L}\bfvg (\bfvx)ds, \label{eqn:semigroupgle}
\end{eqnarray}
\noindent which can be written in terms of the observables:
\begin{eqnarray}
    \label{eqn:observablegle}
 \frac{\partial}{\partial t} \bfvg(\bfvx, t) =\bfvM(\bfvg(\bfvx, t)) +  \bfvF(\bfvx,t) - \int_0^t \bfvK(\bfvg(\bfvx, t-s),s) ds,
\end{eqnarray}
\noindent The specific selection of the projection operators will be discussed below. The above equation is the Generalized Langevin Equation (GLE), which contains a Markov transition term, orthogonal dynamics and a memory kernel that are defined as:

\begin{subequations}
\begin{eqnarray}
\bfvM(\bfvg(\bfvx, t)) &:=&   e^{t\mathcal{L}} P\mathcal{L} \bfvg(\bfvx) 
\label{eqn:GLEmarkov}\\
\bfvF(\bfvx,t)&:=&e^{tQ\mathcal{L}}Q\mathcal{L} \bfvg(\bfvx), \\
\bfvK(\bfvg(\bfvx, t-s),s) &:=& - e^{(t-s)\mathcal{L}}P\mathcal{L}e^{sQ\mathcal{L} }Q\mathcal{L}\bfvg (\bfvx) \nonumber 
\label{eqn:nonlinearGFD}
\end{eqnarray}
\end{subequations}
\noindent with the orthogonality condition $PF(\bfv{x},t)=0$.These three components represent the key construct of Mori-Zwanzig formalism and the GLE is exact in describing the dynamics of observables $\bfvg$. Eq.~\eqref{eqn:nonlinearGFD} is also referred to as the nonlinear Generalized Fluctuation Dissipation relation \citep{zwanzig1973nonlinear}. Note that here we use a negative sign in front of the memory term following the convention in \citet{mori1965transport,zwanzig1973nonlinear}. 





We remark that Eq.~\eqref{eqn:observablegle} is the general form of the MZ formulation and is not specific to any projection operator. In practice, the choice of projection operator is central for constructing the MZ formalism, because the functional form of the components may vary drastically for different projection operators. Here, we give some examples of projections operators that have been employed in the literature for constructing the MZ formalism.

\begin{itemize}
    \item In Mori's formulation \citep{mori1965transport}, the projection operator relies on inner product defined as:
\begin{equation}
\l\langle f , g\r\rangle := \int f\l( \bfvx\r) g\l(  \bfvx\r) \dd \mu\l(\bfvx\r), f,g \in L^2(\mu) \label{eq:ctsInnerProductGeneric}
\end{equation}
\noindent where $f$ and $g$ are $L^2$-integrable functions and $\bfvx$ is drawn from the probability distribution $\mu$. In this work, we adopt a stationary measure for $\bfvx$. With the inner product, the Mori's projection operator \citep{mori1965transport}, or the finite rank projection operator, can be defined onto the span of a set of linearly independent basis functions $g_i(\bfvx)$, $i \in 1,...,D$:
\begin{equation}
    \label{eqn:linearprojectionoperator}
     P f \l ( \bfvg(\bfvx) \r) := \sum_{i,j=1}^{D} \langle f,g_i \rangle \l [\bfvC_0^{-1} \r]_{i,j}g_j(\bfvx),
\end{equation}
\noindent where $\bfvC_0^{-1}$ is the inverse of the covariance matrix $\l[C_0\r]_{i,j} = \langle g_i,g_j \rangle$, $i,j \in 1,...,D$. For a special set of orthonormal basis functions $h_i(\bfvx)$, the covariance matrix becomes identity matrix so that the projection operator can be simplified:
\begin{equation}
    \label{eqn:simplinearprojectionoperator}
     P f \l ( \bfv{h}(\bfvx) \r) := \sum_{i=1}^{D} \langle f,h_i \rangle h_i(\bfvx).
\end{equation}
\noindent In general, one can use the Gram-Schmidt (G-S) procedure to identify the set of orthonornal functions $h_i(x)$ from $g_i(x)$.

    \item In Zwanzig's formulation \citep{zwanzig1973nonlinear}, the observables are chosen to be a sub-set of the variables $\bfvg(\bfvx) = \hat{\bfvx}$ and the projection operator is defined using direct marginalization of the un-resolved variables. If the probability distribution $\mu$ for phase-space variable $\bfvx$ is written for resolved/unresolved variables as $\rho(\hat{\bfvx}, \Tilde{\bfvx})$, the projection operator is then defined as:
\begin{equation}
    \label{eqn:nonlinearprojectionoperator}
     P f \l ( \hat{\bfvx} \r) := \frac{\int f(\hat{\bfvx},\Tilde{\bfvx}) \rho(\hat{\bfvx}, \Tilde{\bfvx})d\Tilde{\bfvx}}{\int\rho(\hat{\bfvx}, \Tilde{\bfvx})d\Tilde{\bfvx}}.
\end{equation}    
\noindent The resulting function $Pf$ is generally nonlinear, so this projection operator is also termed as nonlinear projection \citep{chorin2002optimal} or infinite rank projection \citep{falkena2019derivation}. This nonlinear projection operator has been adopted in \citet{parish2017dynamic,parish2017non} to construct MZ-based models for turbulence, where the initial conditions were assumed to be fully-resolved (i.e, $\Tilde{\bfvx}=0$ at $t=0$) and the unresolved observables were assumed to remain centered at 0 and delta distributed.

\item A recently proposed Wiener projection is used to link the Nonlinear Auto-Regressive Moving Average with eXogenous input (NARMAX) to MZ formalism \citep{lin2019data}, where the basis functions $\bfvg$ also embed information from past history. Let $f_n$ and $g_n$ be two discrete-time zero mean wide-dense stationary processes, where subscript $n$ denotes the index of time steps, then the Wiener projection operator can be written as:
\begin{equation}
    \label{eqn:wienerprojectionoperator}
     P f_n := \sum_{i=1}^{n} h_i g_{n-i},
\end{equation}  
\noindent where the sequence $h_i$ is the Wiener filter.
\end{itemize}

In this work, we focus on the Mori's finite rank projection operator and the corresponding constructed MZ kernels, which is the foundation of the data-driven algorithms proposed in \citet{lin2021data}.

\subsubsection{A Discrete-time Mori-Zwanzig Formalism}

Even though the dynamical system discussed above is formulated in continuous-time, it is common that in high-resolution simulations or experimental measurements, the outputs are discrete-time snapshots, where the temporal derivative is not readily available. For completeness, in this section, we introduce the discrete formulation of dynamical system and corresponding MZ formulation following \citet{lin2019data}

We write the dynamical equation for the full set of discrete solution vector $\bfvPhi({n\Delta}) \in \mathbb{R}^N$as:
\begin{equation}
 \label{eqn:discreteode}
     \bfvPhi((n+1)\Delta )=S_\Delta(\bfvPhi(n\Delta )), \bfvPhi(0)=\bfvx,
 \end{equation}

\noindent where $n$ and $\Delta$ are the time step and time interval of the discrete-time snapshots, respectively. Similar to the continuous time derivation, we define the observables as $\bfvg(\bfvPhi(n\Delta))\in \mathbb{R}^D$, where the components $g_i$ are functions in $L^2(\mu)$ that map the original solution $\bfvPhi$ to a physically observed quantity of interest: $\mathbb{R}^N \rightarrow \mathbb{R}$. For simplicity, we use $\bfvg(n\Delta)$ to denote $\bfvg(\bfvPhi(n\Delta))$. To describe the evolution of the observables $\bfvg$, we introduce discrete-time Koopman operator $\mathcal{K}_\Delta$ that satisfies $\l[\mathcal{K}_\Delta\bfvg\r](\bfvPhi)=(\bfvg\circ S_\Delta )(\bfvPhi)$, where the symbol $\circ$ is used to denote composite functions. By operating the Koopman operator on functions $\bfvg$ and applying to the solution variable at the current state, we can obtain the observables at the future time step. We apply the Koopman operator $n+1$ times and derive the evolution of  observables:

\begin{equation}
    \label{eqn:koopmanoperator}
   \l[\mathcal{K}_\Delta^{n+1}\bfvg \r] (\bfvx) =\l[\mathcal{K}_\Delta^{n}(\bfvg \circ S_\Delta)\r] (\bfvx) =\l [\mathcal{K}_\Delta^n \bfvg\r](\bfvPhi(\Delta)) = \bfvg(\bfvPhi((n+1)\Delta)) = \bfvg((n+1)\Delta).
\end{equation}

\noindent With a given projection operator $P$ on Hilbert space $\mathcal{H}$ and its complement $Q=I-P$, we can then write the Dyson identity for the Koopman operator \citep{lin2019data}:

\begin{equation}
    \label{eqn:dysondiscrete}
\mathcal{K}_\Delta^{n+1} = \sum_{l=0}^{n}\mathcal{K}_\Delta^{n-l}P\mathcal{K}_\Delta(Q\mathcal{K}_\Delta)^l + (Q\mathcal{K}_\Delta)^{n+1}.
\end{equation}

\noindent We substitute Eq.~\eqref{eqn:dysondiscrete} in Eq.~\eqref{eqn:koopmanoperator} and arrive at the evolutionary equation for observables $\bfvg$:

\begin{eqnarray}
    \label{eqn:GLEdiscrete}
    \bfvg((n+1)\Delta) &=& \l [\mathcal{K}_\Delta^{n+1}\bfvg\r](\bfvx) = \sum_{l=0}^{n}\l [\mathcal{K}_\Delta^{n-l}P\mathcal{K}_\Delta(Q\mathcal{K}_\Delta)^l\bfvg\r](\bfvx) + \l [(Q\mathcal{K}_\Delta)^{n+1}\bfvg\r](\bfvx) \nonumber \\
    &=&\l[\mathcal{K}_\Delta^{n}P\mathcal{K}_\Delta\bfvg\r](\bfvx)+ \sum_{l=1}^{n}\l[\mathcal{K}_\Delta^{n-l}P\mathcal{K}_\Delta(Q\mathcal{K}_\Delta)^l\bfvg\r](\bfvx) + \l[(Q\mathcal{K}_\Delta)^{n+1}\bfvg\r](\bfvx) \nonumber \\
    &=& \bfvO^{(0)}_\Delta (\bfvg(n\Delta)) + \sum_{l=1}^{n} \bfvO^{(l)}_\Delta( \bfvg((n-l)\Delta)) + \bfvW_{n+1}(\bfvx).
\end{eqnarray}

The above equation is the discrete-time GLE and can be understood as the discrete counterpart of Eq.~\eqref{eqn:semigroupgle}. The corresponding Mori-Zwanzig operators can then be derived for the three components in Eq.~\eqref{eqn:GLEdiscrete}:

\begin{eqnarray}
 \bfvO^{(0)}_\Delta ( \bfvg(n\Delta)) &=& \l[\mathcal{K}_\Delta^{n}P\mathcal{K}_\Delta\bfvg\r](\bfvx) 
 , \\
 \bfvW_{n+1}(\bfvx) &=&\l[(Q\mathcal{K}_\Delta)^{n+1}\bfvg\r](\bfvx), \\
 \bfvO^{(l)}_\Delta (\bfvg((n-l)\Delta)) &=& \l[\mathcal{K}_\Delta^{n-l}P\mathcal{K}_\Delta(Q\mathcal{K}_\Delta)^l\bfvg\r](\bfvx),
\end{eqnarray}
\noindent with the orthogonality condition $P\bfvW_{n+1}=0,$ $\forall$ $n \in \mathcal{N}$. Eq.~\eqref{eqn:GLEdiscrete} is also general and the specific forms of the MZ operators depend on the choice of projection operator.

\subsection{Reduced-order Construction of Navier-Stokes Equation Based on MZ Formalism}
\label{sec:NS-ROM}
In the previous section, we derived and discussed the key components of MZ formalism and their dependence on the projection operator. In this section, we first demonstrate their application on Navier-Stokes turbulence modeling and establish the link between MZ formulation with the nonlinear projection operators and classical turbulence modeling approach, i.e. LES. Due to the challenges in extracting the properties of the corresponding nonlinear MZ operators, we propose the MZ formulation based on Mori's finite rank projection operator as a generalization of Koopman learning framework, which greatly simplifies the learning task and allows us to quantitatively address the assumptions in modeling.

\subsubsection{LES and MZ Formulation Based on Nonlinear Projection operator}
Consider the three-dimensional discretized velocity field in a fully-resolved numerical simulation $v_i(t, n_x, n_y, n_z)$, where $i \in 1,2,3$ is the direction of velocity and $n_x, n_y, n_z \in 1,..,N_x, N_y, N_z$ the spatial coordinates of the velocity field. We can stack the discretized solution of the velocity field at time t in to a $N\times1$ vector $\bfvv(t) \in \mathbb{R}^N$, $N=3\times N_x \times N_y \times N_z$, and write the discretized incompressible Navier-Stokes equation with any given numerical scheme into a general form ODE, which follows Eq.~\eqref{eqn:ode}:
\begin{equation}
\label{eqn:nsode}
\frac{d\bfvv(t)}{dt}=R(\bfvv(t)),
\end{equation}
\noindent where $R$ are nonlinear functions that can be viewed as the spatially-discretized form of the right hand side of the Navier-Stokes equation given a numerical scheme. 

For the spatially discretized N-S equation, we can then derive the discrete-time formulation for the temporally-discretized velocity vector $\bfvv(n\Delta)$ and arrive at Eq.~\eqref{eqn:discreteode}. Here, $S_\Delta$ encodes information of the temporal scheme, for example, $S_\Delta(\bfvv) = I + \Delta \cdot R(\bfvv)$ for the Euler method.

 Fully resolving the dynamics of the Navier-Stokes equations requires prohibitively large amounts of computational resources due to the wide range of scales of practically relevant problems. In the classic approach of reduced-order modeling for turbulence, the velocity field is coarse-grained by applying a spatial filter to reduce the range of scales that need to be resolved. Here, we denote the solution vector of the filtered discretized velocity field as $\overline{\bfvv}(t):=[\overline{v_i}(t,n_x,n_y,n_z)]^T$, $n_x,n_y,n_z \in 1,...,N_{x,c},N_{y,c},N_{y,c}$, where the overline denotes the spatial filtering. Commonly used spatial filters include Gaussian filter, box filter, and spectral filter.  The size of the computational mesh required to fully resolve $\overline{\bfvv}$ is significantly reduced because of the reduced range of scales, so that $N_{x,c} \ll N_{x}$, $N_{y,c} \ll N_{y}$, $N_{z,c} \ll N_{z}$ and the resolved solution vector $\overline{\bfvv} \in \mathbb{R}^D$ has a reduced dimension $D=3\times N_{x,c}\times N_{y,c} \times N_{z,c} \ll N$. As a result of spatially filtering the nonlinear NS equation, high-order moments emerge and the system for the resolved variable $\overline{\bfvv}$ is not closed. Dynamical equations for the filtered velocities can be written as:

\begin{equation}
    \frac{d}{dt}\overline{\bfvv}(t)=\overline{R}(\overline{\bfvv}(t)) + \tau_{sgs}(\bfvv(t)),
\label{eqn:nsresolvediscrete}
\end{equation}
\noindent where $\overline{R}(\overline{\bfvv}(t))$ takes the same form as the original NS equations but numerically discretized on a coarser grid and is fully closed, while $\tau_{sgs}(\bfvv(t))$ denotes the unclosed sub-grid scale contributions. Transport equations for the higher-order moments can also be derived, such as equations for the sub-grid stress (SGS) in LES, but they depend on even higher-order unclosed terms. In practice, the resulting infinite dimensional system is truncated to include only the resolved variable $\hat{\bfvPhi}=\overline{\bfvv}$, and a sub-grid model is used to compensate for the effects from the unresolved moments/scales, which is usually Markovian: $\tau_{sgs}(\overline{\bfvv}(t))$.

In the framework of MZ formalism, we can write evolution equation of the reduced-order variable $\overline{\bfvv}$ as a GLE that consists of a Markov term, a memory term and orthogonal dynamics:

\begin{eqnarray}
\label{eqn:nsmarkov}
\frac{d}{dt}\overline{\bfvv}(t) &=& M(\overline{\bfvv}(t)) + F(t) - \int_0^tK\l (\overline{\bfvv }(t-s),s\r)ds.
\end{eqnarray}
\noindent

We remark that there may exist a projection operator that one can apply to the filter NS equations and establish the connection between the terms in Eq.~\ref{eqn:nsresolvediscrete} and the terms in MZ formulation Eq.~\ref{eqn:nsmarkov}, but it is challenging to perform quantitative analyses of the sub-grid model in the MZ framework. In this section, we hope to shed some light on turbulence modeling from the perspective of MZ formalism. The traditional sub-grid scales models that are based on assumptions of scale similarity, flow homogeneity, etc. have been shown to be inadequate in many complex turbulent problems, such as transitional flow, flow with separation, etc. This may originate from the lack of consideration of the mathematical properties that SGS model needs to satisfy with regard to the projection operator, memory kernel, and orthogonal dynamics. Alternatively, we can develop turbulence models from the MZ framework, where the complex transient dynamics can be naturally incorporated into the memory kernel and orthogonal dynamics. This will alleviate the need for making assumptions that relate the small-scale fluctuations to velocities at the resolved scale $\overline{\bfvv}$. On the other hand, it is generally difficult to accurately model the Markov term, memory kernel, and orthogonal dynamics due to limited knowledge of their properties and the challenge in extracting MZ operators using data/observations.

We also remark that the difficulty of extracting the memory kernel and orthogonal dynamics originates from two aspects: nonlinearity of the Navier-Stokes equations and that of the projection operator. In the next section, we solve this issue by introducing Mori's finite rank projection operator and discuss its relation to the Koopman learning framework, which lays the foundation for the extraction of MZ kernels using a data-driven algorithm. We also remark that we only discuss Smagorinsky-type sub-grid models \citep{smagorinsky1963general} in Eq.~ \eqref{eqn:nsresolvediscrete} for simplicity, while there exists a wide range of models, such as one-equation model \citep{yoshizawa1985statistically}, dynamic model \citep{germano1991dynamic}, etc., in which certain memory effects are incorporated, even though indirectly.

\subsubsection{MZ Formulation Based on Mori's Projection Operator}
 
In the Koopman learning framework \citep{koopman1931hamiltonian}, the same dynamical system in Eq.~\eqref{eqn:ode} can be characterized by a collection of observables $\bfvg$, which are functions of the physical-space variables $\bfvPhi$. The system can then be cast from a finite-dimensional system of nonlinear ODEs describing the physical variables to an infinite-dimensional system of linear ODEs that describes all possible observables. In Koopman's formulation, the observables evolve on an infinite-dimensional Hilbert space $\mathcal{H}$, which is composed of all possible observables.

In the Koopman framework, deriving a closed-form solution is equivalent to identifying a set of observables whose dynamics are invariant in a subspace which is linearly spanned by the set of the observables. In general, it is very challenging to identify the finite set of observables that close the dynamics and one has to resort to approximation methods to close the dynamics. Naturally, we can leverage the Mori-Zwanzig formalism and the inner product equipped in the Hilbert space to construct the dynamical equations for the finite set of observables. \citet{lin2021data} showed that by using Mori's finite rank projection operator (Eq.~\eqref{eqn:linearprojectionoperator}), which depends on the inner product of two functions, both the Koopman \citep{koopman1932dynamical} and MZ formulations operate in a shared Hilbert space. The advantage of using Mori's projection operator is that the projected low-dimensional functions are linear, which significantly simplifies the derivation/learning of the MZ kernels. This is in contrast to the MZ construction based on nonlinear projection operators as discussed in the previous section. Following \citet{lin2021data}, the MZ formulations when using the Mori's projection operator with linearly independent basis functions $\bfvg$ can be written as:

\begin{equation}
\frac{\dd}{\dd t} \bfvg\l(t\r) = \bfvM \cdot \bfvg\l(t\r) - \int_{0}^t \bfvK\l(t-s\r)\cdot \bfvg\l(s\r)\dd s + \bfvF\l(t\r),\label{eqn:moriGLE}
\end{equation}

\noindent where $\bfvM$ and $\bfvK$ are $D \times D$ matrices. Similarly, the discrete counterpart of MZ formulation based on Mori's projection operator for the discrete-time observable $\bfvg(n\Delta) = \bfvg(\bfvPhi(n\Delta))$ can be written as \citep{lin2021data}:

\begin{eqnarray}
    \bfvg((n+1)\Delta)  &=&   \bfvO_\Delta^{\l(0\r)} \cdot  \bfvg(n\Delta) + \sum_{\ell=1}^{n} \bfvO_\Delta^{\l(\ell\r)} \cdot  \bfvg((n-l)\Delta) + \bfvW_{n+1} \nonumber\\
    &=& \sum_{\ell=0}^{n} \bfvO_\Delta^{\l(\ell\r)} \cdot  \bfvg((n-l)\Delta) + \bfvW_{n+1}  \label{eqn:dmoriGLE}
\end{eqnarray}

\noindent Note that in the discrete form, $\bfvO_\Delta^{\l(0\r)}$ is the Markov operator, $\bfvO_\Delta^{\l(l\r)}$, $l \in 1,2,3..$ are the memory kernels, $\bfvW_{n+1}$ is the orthogonal dynamics, and $\Delta$ represents the discrete time step \citep{lin2021data}. There is also a switch of sign in the memory kernel between the two types of formulation in equations \ref{eqn:moriGLE} and \ref{eqn:dmoriGLE}, which is because of the conventions in continuous and discrete formulations. In the rest of the paper, the MZ formulation will be mainly discussed using the discrete form, due to the fact that it simplifies the calculation by replacing integral with summation and temporal gradient with numerical values.

We remark that Eq.~\eqref{eqn:moriGLE} (or Eq.~\eqref{eqn:dmoriGLE} for discrete-time formulation) is a special case of the general MZ formulation (Eq.~\eqref{eqn:semigroupgle} and \eqref{eqn:GLEdiscrete}) where the projection operator is chosen to be the finite rank projection operator. The basis functions $\bfvg$ are not limited to the original physical-space variables, $\bfvg(\bfvPhi)=\bfvPhi$, but can be any linearly independent functions of $\bfvPhi$. Naturally, we can construct a MZ formulation for the Navier-Stokes equation using the Mori's projection operator. The form of the MZ formulation follows that in Eq.~\eqref{eqn:moriGLE}.

So far, we have presented two different Mori-Zwanzig formulations for the N-S equation (Eqs.~\eqref{eqn:moriGLE}/\eqref{eqn:dmoriGLE} and \eqref{eqn:nsmarkov}/\eqref{eqn:GLEdiscrete}), and their differences can be understood from two different perspectives. {\em Firstly, the projection operators are different}: Eq.~\eqref{eqn:nsmarkov} employs a projection operator based on truncation, which results in nonlinear Markov operator and memory kernels, while Eq.~\eqref{eqn:moriGLE} is based on Mori's projection operator, which results in linear Markov and memory kernels. We have discussed the relation between Eq.~\eqref{eqn:nsmarkov} and LES modeling and difficulties of deriving/extracting corresponding memory kernels and orthogonal dynamics in the previous section. This difficulty can be alleviated using the linear MZ formulation in Eq.~\eqref{eqn:moriGLE}, but there is no direct link to classical LES modeling. {\em Secondly, the solution vectors are different}: Eq.~\eqref{eqn:nsmarkov} describes the evolution of physical-space variables (the filtered velocity field in the context of LES), while Eq.~\eqref{eqn:moriGLE} describes the observables $\bfvg$ which can be nonlinear functions of physical-space variables. The observables can also include the physical-space variables themselves in their set. 

Given the simplicity of the linear Markov operator and memory kernels when using Mori's projection operator, we will show that the Mori-Zwanzig operators can be extracted in a relatively straightforward manner using this formulation, unlike the one based on a nonlinear projection operator. We will also continue the discussions using the discrete-time formulation because: (i) observations and simulation results are usually discrete and fully resolved temporal gradients are usually not readily available, and (ii) discrete-time formulation can avoid the errors induced by numerically integrating the continuous-time counterpart.

\subsection{The Evolution of Time Correlation Matrix}

In this section, we derive the evolution equation of the time correlation matrix $\bfvC$, which is the foundation of the learning algorithm. For Mori's projection operator, we choose to use the initial condition $\bfvg(0)$ as the basis of the projected linear subspace. We then apply an inner product $\l\langle \cdot \bfvg(0) \r\rangle$ to Eq.~\eqref{eqn:dmoriGLE} and obtain an evolutionary equation for the two-time correlation function $\bfvC(n\Delta) = \l\langle \bfvg(n\Delta),\bfvg(0)^T  \r\rangle$:

\begin{align}
\bfvC(n\Delta) & = \l\langle \bfvg(n\Delta),\bfvg(0)^T \r\rangle,  \\
\bfvC((n+1)\Delta) &= \bfvO_\Delta^{\l(0\r)} \cdot  \bfvC(n\Delta) + \sum_{\ell=1}^{n} \bfvO_\Delta^{\l(\ell\r)} \cdot  \bfvC((n-l)\Delta)=\sum_{\ell=0}^{n} \bfvO_\Delta^{\l(\ell\r)} \cdot  \bfvC((n-l)\Delta). \label{eqn:twotimecorrelation}
\end{align}

\noindent Note that with this procedure, we exploit the orthogonality between the basis function $\bfvg(0)$ and the noise $\bfvW_{n+1}$ to remove the complex orthogonal dynamics in the evolution equation of $\bfvC$. The resulting formula (Eq.~\eqref{eqn:twotimecorrelation}) builds the foundation for the data-driven learning algorithm.

\subsection{Generalized Fluctuation Dissipation Relation}

The relation between the memory kernel and the orthogonal dynamics, i.e. Eq.~\eqref{eqn:nonlinearGFD}, is commonly referred to as the Generalized Fluctuation Dissipation relation (GFD). There exist different interpretations of this relation but, in general, it imposes a structural relation between the memory kernel and orthogonal dynamics. Thus, these operators can not be approximated using models in an arbitrary manner. This relation has been used to estimate the memory kernel when a model for orthogonal dynamics is proposed \citep{gouasmi2017priori}.

When constructing the MZ formulation using Mori's projection operator, a specific form of the GFD can be derived for the two-time correlation matrix, if the Liouville operator $\mathcal{L}$ is anti-self-adjoint with respect to the chosen inner product:

\begin{equation}
    \label{eqn:antiselfadjoint}
\l \langle f, \mathcal{L}h\r\rangle = -\l \langle \mathcal{L}f, h\r\rangle,
\end{equation}
\noindent for any functions $f$ and $h$ of the physical-space variable $\bfvPhi$. Note that the anti-self-adjoint property depends on the choice of the inner product. \citet{lin2021data} showed that the Liouville operator of a dynamical system is anti-self-adjoint if the inner product is defined as the temporally averaged value of the product of the test functions evaluated on a long trajectory, provided the observables are bounded along the trajectory. The specific GFD for the discrete-time MZ formulation with Mori's projection operator is \citep{lin2021data}:

\begin{equation}
    \bfvO_\Delta^{(l)} =- \l\langle \bfvW_{l+1} \cdot \bfvW_1 \r\rangle \bfvC^{-1}(-\Delta), \hspace{0.5cm} \forall l \in 1,2,3...,
    \label{eqn:GFD}
\end{equation}
\noindent where $\bfvC(-\Delta)=\bfvC^T(\Delta)$. This non-trivial relation should be satisfied if the kernels are correctly extracted from the data and will be verified using numerical data.

\section{Data Driven Learning of the MZ Operators and Orthogonal Dynamics}
\label{sec:learning}

In this section, we briefly describe the learning algorithm proposed in \citet{lin2021data}. The learning procedure is based on Eq.~\eqref{eqn:twotimecorrelation} and starts by calculating the two-time correlation matrix $\bfvC(n\Delta) = \l\langle \bfvg(n\Delta),\bfvg(0)^T  \r\rangle$. As mentioned in previous section, the evaluation of the inner product requires taking expectation value against the stationary distribution $d\mu$ or temporally and uniformly sampling/averaging along a long trajectory. Given a long and evenly spaced trajectory of physical space variable (velocity field) from the fully resolved Direct Numerical Simulation $\bfvPhi(n\Delta), n \in 0,1...N_t-1$, the two-time correlation matrix $\bfvC(n\Delta)$ can be calculated as:

\begin{align}
\bfvC\l(n\Delta \r) = \frac{1}{N_t-n} \sum_{i=0}^{N_t-n-1} \bfvg (\bfvPhi\l(\l(n+i\r) \Delta \r)) \cdot \bfvg^T ( \bfvPhi\l( i\Delta \r)) , \label{eq:disC} 
\end{align}

\noindent where $N_t$ is the total number of snapshots and $N_t\Delta \gg T_{l}$, where $T_l$ is the integral time scale of turbulence. It is also beneficial to implement known symmetries of the physical system for data augmentation, which can further improve the accuracy in extracting Mori-Zwanzig operators without generating more data. For isotropic turbulence with triply-periodic boundaries, periodicity and rotational symmetries are satisfied and can be used to facilitate data augmentation. This can be implemented in the calculation of the two-time correlation matrix. Suppose there exist $N_s$ symmetric representations of the same physical-space variable $\mathcal{S}_{n_s}(\bfvPhi), n_s = 1,...N_s$, where $\mathcal{S}_{n_s}$ is the symmetry operator that preserve the statistics of the original physical-space variables $\bfvPhi$. One of such operator could be rotating the velocity field such that $u_1 \rightarrow u_2, u_2 \rightarrow u_3, u_3 \rightarrow u_1$, and the dynamics are the same for all the symmetric representations. Naturally, Eq.~\eqref{eq:disC} can then be modified to:

\begin{align}
\bfvC\l(n\Delta \r) = \frac{1}{N_s} \frac{1}{N_t-n} \sum_{n_s=1}^{N_s}\sum_{i=0}^{N_t-n-1} \bfvg (\mathcal{S}_{n_s}(\bfvPhi\l(\l(n+i\r) \Delta \r))) \cdot \bfvg^T ( \mathcal{S}_{n_s}(\bfvPhi\l( i\Delta \r))). \label{eq:symmetricdisC} 
\end{align}

After the calculation of two-time correlation matrix $\bfvC(n\Delta)$, we can set $n=0$ in Eq.~\eqref{eqn:twotimecorrelation} to obtain the Markov operator $\bfvO_\Delta^{(0)}$:

\begin{equation}
    \label{eqn:markov}
    \bfvO_\Delta^{(0)} = \bfvC(\Delta) \cdot \bfvC^{-1}(0).
\end{equation}

\noindent We can then recursively solve for the memory kernel $\bfvO_\Delta^{(n)}$ using two-time correlation matrix $\bfvC(n\Delta)$, Eq.~\eqref{eqn:twotimecorrelation} and previously solved low-order $\bfvO_\Delta^{(l)}, l < n$:

\begin{equation}
    \label{eqn:memorykernel}
    \bfvO_\Delta^{(n)} = (\bfvC((n+1)\Delta) - \sum_{l=0}^{n-1} \bfvO_\Delta^{(l)}  \bfvC((n-l)\Delta) )\cdot \bfvC^{-1}(0).
\end{equation}

After we obtain memory kernel, the orthogonal dynamics can then be extracted for each section in the trajectory using Eq.~\eqref{eqn:dmoriGLE}:

\begin{equation}
    \label{eqn:noise}
    \bfvW_{n+1}^{(i)} = \bfvg(\bfvPhi((i+n+1)\Delta)) - \sum_{l=0}^{n-1} \bfvO_\Delta^{(l)}  \bfvg(\bfvPhi((i+n-l)\Delta)).
\end{equation}

\noindent After obtaining $\bfvW_{n+1}^{(i)}$, the properties of the orthogonal dynamics can be studied, such as the two-time correlation $\l\langle \bfvW_{n+1}, \bfvW_1\r\rangle$, etc.

\section{Ground Truth Turbulence Data}
\label{sec:dnsdata}

The "Ground-Truth" data is generated from the Eulerian DNS solution of the incompressible Navier-Stokes equations:

\begin{equation}
\frac{\partial v_i}{\partial t} +\frac{\partial v_iv_j}{\partial x_j} = -\frac{\partial p}{\partial x_i} + \nu \frac{\partial^2 v_i}{\partial x_j x_j},
\end{equation}
where pressure $p$ is obtained by solving the Poisson equation and $\nu$ is the Kinematic viscosity. The isotropic turbulence is generated on a $128^3$ grid using the pseudo-spectral method. A large scale forcing term is applied to prevent turbulence from decaying. Time advancement is achieved through Adam-Bashforth-Moulton method. The Taylor Reynolds number when the turbulence reaches a statistically-steady state is $\approx 100$. See \citet{petersen2010forcing,daniel2018reaction} for more details on the numerical method. 


After the turbulent flow becomes fully-developed, the 3D snapshots of the flow field are stored in consecutive time steps to generate a long trajectory of turbulence data. The total length of the trajectory is approximately $3000 T_l$, where $T_l$ is the integral time scale. 
We then apply the post-processing procedure to obtain the low-dimensional coarse-grained observables. The choices of observables could significantly affect the properties of the Markov operator, memory kernels, and the noise, therefore in this study, we select several sets of observables that are closely related to the canonical turbulence modeling approach (LES) and turbulence theory. To summarize, the following procedures are used in this work to obtain the observables: 1) similar to LES, we first apply spatial filters to the velocity field at each time step with a wide range of filter sizes $l_\Delta$ to obtain the filtered velocity $\overline{v_i}$,  2) various types of observables such as filtered velocity, pressure, sub-grid stress, kinetic energy, etc.~, are then computed from the spatially-filtered field, 3) the selected observables on the fine mesh ($128\times 128 \times 128$) are then uniformly sampled onto a coarse mesh ($4 \times 4 \times 4$), and 4) the resulting observables at each time step are stacked into a single vector. Following these coarse-graining steps, a dimension reduction from $3\times128^3$ physical-space variables to $n_{obs}\times4^3$ observables can be achieved, where $n_{obs}$ is total number of function/variable types. We include the filtered velocities in all sets of observables. In this work, we consider the following four sets of observables for extracting MZ kernels.

\begin{itemize}
    \item Observable set 1 ($n_{obs}$ = 3): $\overline{v_1}$, $\overline{v_2}$, $\overline{v_3}$
    \item Observable set 2 ($n_{obs}$ = 15): $\overline{v_1}$, $\overline{v_2}$, $\overline{v_3}$, $\overline{v_1}\,\overline{v_1}$, $\overline{v_2}\,\overline{v_2}$, $\overline{v_3}\,\overline{v_3}$, $\overline{v_1}\,\overline{v_2}$, $\overline{v_1}\,\overline{v_3}$, $\overline{v_2}\,\overline{v_3}$, $\overline{v_1v_1}-\overline{v_1}\,\overline{v_1}$, $\overline{v_2v_2}-\overline{v_2}\,\overline{v_2}$, $\overline{v_3v_3}-\overline{v_3}\,\overline{v_3}$, $\overline{v_1v_2}-\overline{v_1}\,\overline{v_2}$, $\overline{v_1v_3}-\overline{v_1}\,\overline{v_3}$, $\overline{v_2v_3}-\overline{v_2}\,\overline{v_3}$,
    
    \item Observable set 3 ($n_{obs}$ = 14): $\overline{v_1}$, $\overline{v_2}$, $\overline{v_3}$,  $\overline{v_1}\,\overline{v_1}+\overline{v_2}\,\overline{v_2}+\overline{v_3}\,\overline{v_3}$, $S_{11}$, $S_{22}$, $S_{12}$, $S_{13}$, $S_{23}$, $W_{12}$, $W_{13}$, $W_{23}$, $S_{ij}S_{ij}$, $W_{ij}W_{ij}$, where $S_{ij} = \frac{1}{2}(\frac{\partial \overline{v_i}}{\partial x_j} + \frac{\partial \overline{v_j}}{\partial x_i})$, $W_{ij} = \frac{1}{2}(\frac{\partial \overline{v_i}}{\partial x_j} - \frac{\partial \overline{v_j}}{\partial x_i})$
    \item Observable set 4 ($n_{obs}$ = 15): $\overline{v_1}$, $\overline{v_2}$, $\overline{v_3}$, $\frac{\partial \overline{v_1v_1}}{\partial x_1}$, $\frac{\partial \overline{v_2v_2}}{\partial x_2}$, $\frac{\partial \overline{v_3v_3}}{\partial x_3}$, $\frac{\partial \overline{v_1v_2}}{\partial x_1}$, $\frac{\partial \overline{v_1v_2}}{\partial x_2}$, $\frac{\partial \overline{v_1v_3}}{\partial x_1}$, $\frac{\partial \overline{v_1v_3}}{\partial x_3}$, $\frac{\partial \overline{v_2v_3}}{\partial x_2}$, $\frac{\partial \overline{v_2v_3}}{\partial x_3}$, $\frac{\partial \overline{p}}{\partial x_1}$, $\frac{\partial \overline{p}}{\partial x_2}$, $\frac{\partial \overline{p}}{\partial x_3}$

\end{itemize}

\noindent Note that in observable set 3, not all components of the strain rate tensor $S_{ij}$ are chosen as basis functions for observables because there exists a linear dependence of the diagonal components from the incompressibility condition. In addition to the above-mentioned functions for each set of the observables, a constant function $g_0=1$ is added to every set of observables, making the total number of observables $n_{obs}\times4^3+1$. 


\section{Results}
\label{sec:results}
In this section, we present the results and properties of the extracted MZ kernels. We first address the statistical convergence of the learned MZ operators from "Ground-Truth" data. The non-trivial GFD relation between the memory kernels and orthogonal dynamics is also verified. The properties of the Markov operator, memory kernel and noise and their dependence on different choices of observables are then analyzed. Lastly, we conduct numerical experiments using the extracted kernels to investigate the effects of memory kernel on prediction.

\subsection{Statistical Convergence}
\label{sec:convergence}
The statistical convergence of the computed two-time correlation matrix $\bfvC$ and the learned kernel is an important factor in the proposed learning algorithm \citep{lin2021data} and needs to be confirmed in order to reduce the effects of statistical variability on the analysis. The accurate computation of the two-time correlation matrix $\bfvC$ in the ergodic system requires averaging over a long trajectory. This requires a large amount of data samples, which represent the distribution of the stationary system. As described in Section \ref{sec:dnsdata}, we performed fully-resolved simulations to generate a long trajectory of 3D turbulence data and stored total $N_t$ 3D snapshots of velocity fields. In the convergence test, three different sampling methods are used to sample from the database:

\begin{itemize}
    \item Method 1: The convergence test data are randomly sampled from the total $N_t$ snapshots.
    \item Method 2: The total $N_t$ snapshots are first coarsely sampled in time to make sure that any two snapshots are at least one integral timescale apart. An integral timescale $T_l \approx N_{int}dt$ corresponds to approximately $N_t/N_{int}$ snapshots. The convergence test data are then randomly sampled from the $N_t/N_{int}$ snapshots. This procedure ensures that the data are not temporally-correlated and are truly independent samples from the stationary distribution.
    \item Method 3: Similar to Method 2, but the time differences between two snapshots are reduced to half integral timescale. This works as an intermediate case between Method 1 and Method 2.
\end{itemize}

\begin{figure}[hbt]
    \centering
    \includegraphics[width=6in,height=2.5in]{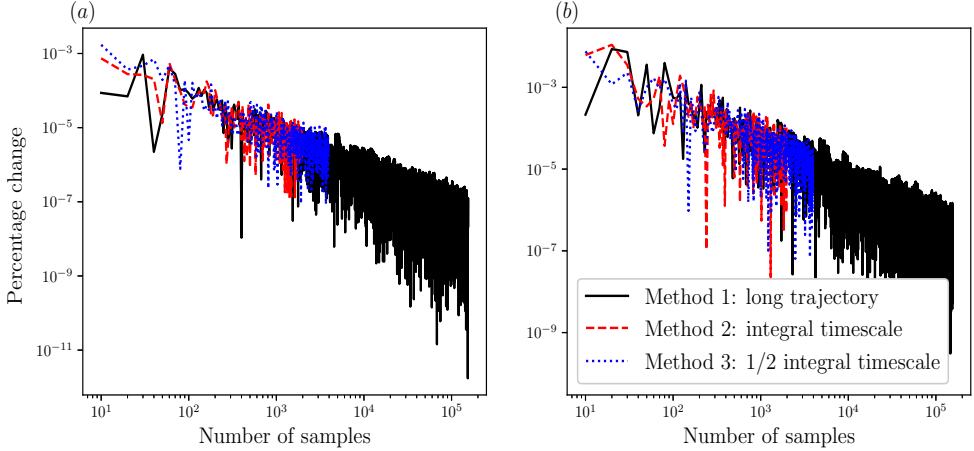}
    \caption{The percentage error of Frobenius norm of the learned (a) Markov operator $\bfvO_\Delta^{\l(0\r)}$ and (b) memory kernel $\bfvO_\Delta^{\l(1\r)}$ as the number of samples increases is shown to verify the statistical convergence. The percentage error is calculated as $\frac{|\Vert\bfvO_\Delta^{\l(\ell\r)}\Vert_{F,n+1}-\Vert\bfvO_\Delta^{\l(\ell\r)}\Vert_{F,n}|}{\Vert\bfvO_\Delta^{\l(\ell\r)}\Vert_{F,n}}$.}
    \label{fig:converge_kernel}
\end{figure}

After obtaining the samples from the database, we apply the post-processing procedure as discussed in Section \ref{sec:dnsdata} to the 3D simulation data in order to generate the vectors of observables $\bfvg$. In the convergence test, the observable set 1 is chosen with the a spatial filtering length $\pi/8$. Data augmentation based periodicity and rotational symmetry is also performed for the statistical convergence test. The discrete time step $\Delta$ is chosen to be $10dt$. Figure \ref{fig:converge_kernel} shows the percentage changes of the Frobenius norm of the learned Markov operator and memory kernels with different number of samples, where the percentage change is calculated using formula: $\frac{|\Vert\bfvO_\Delta^{\l(\ell\r)}\Vert_{F,n+1}-\Vert\bfvO_\Delta^{\l(\ell\r)}\Vert_{F,n}|}{\Vert\bfvO_\Delta^{\l(\ell\r)}\Vert_{F,n}}$. Here, the subscript $F$ denotes the Frobenius norm and $n$ denotes the number of samples used for calculating the Markov operator $\bfvO_\Delta^{(0)}$ and memory kernels $\bfvO_\Delta^{(l)}, l \in 1,2,3...$. The percentage error fluctuates as the number of sample increases, but there exists a converging trend for the upper bound of the fluctuations. By using the largest number of samples from the database, the upper bound of the fluctuations can be reduced to $10^{-7} \sim 10^{-6}$, implying that the final percentage error is less than $10^{-6}$. Further increasing the number of samples only yields negligible improvements in the accuracy of the learned kernels. 
It is also interesting to note that there is no difference in the rate of convergence among the three sampling methods. Considering this, we will use Method 1 for sampling, as it uses the largest amount of data, thus resulting in the smallest error in the learned kernel.

\begin{figure}[hbt]
    \centering
    \includegraphics[width=6in]{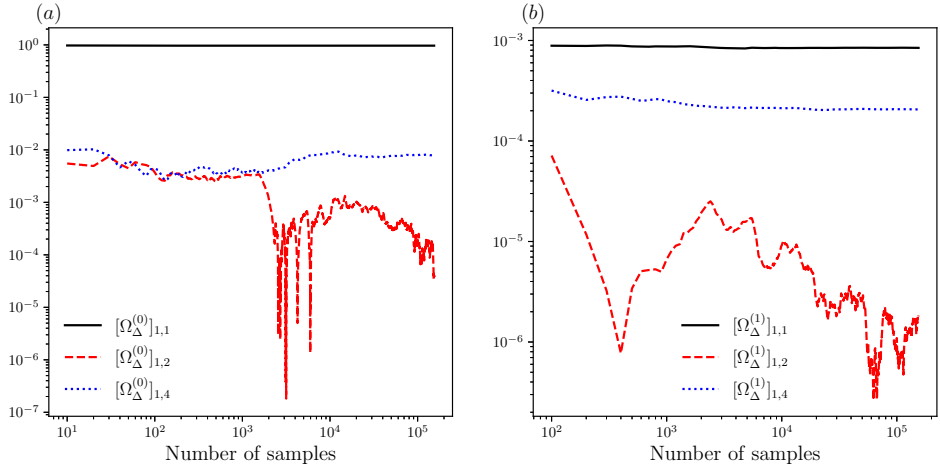}
    \caption{The convergence of components in the learned Markov operator (a) and memory kernel (b) as the number of samples increases.}
    \label{fig:converge_component}
\end{figure}

Considering that for large matrices like the Markov term and memory kernel the convergence of their Frobenius norm does not necessarily translate to the convergence of individual components of the matrices, we provide further test to ensure the full convergence of each component. Figure \ref{fig:converge_component} shows the convergence of the learned individual components of the Markov operator and memory kernels as the number of samples increases. Three components $[\bfvO_{\Delta}^{\l(0\r)}]_{1,1}$, $[\bfvO_{\Delta}^{\l(0\r)}]_{1,2}$, and $[\bfvO_{\Delta}^{\l(0\r)}]_{1,4}$ are shown in figure \ref{fig:converge_component} (a), which corresponds to Markovian contributions from different observables. It can be seen that both $[\bfvO_{\Delta}^{\l(0\r)}]_{1,1}$ (representing contributions from $\overline{v_1}$ to $\overline{v_1}$) and $[\bfvO_{\Delta}^{\l(0\r)}]_{1,4}$ (represents contributions from $\overline{v_1}$ to $\overline{v_1}$ at neighboring points) values achieve convergence when the number of samples is larger than $10^5$. On the other hand, the component $[\bfvO_{\Delta}^{\l(0\r)}]_{1,2}$ (representing contributions from $\overline{v_1}$ to $\overline{v_2}$) shows a decreasing trend as the number of sample increases and reaches almost three orders of magnitude smaller than the dominant components in the matrix. Intuitively, the correlation between velocity components should vanish in an isotropic turbulent field, which implies that $[\bfvO_{\Delta}^{\l(0\r)}]_{1,2}$ is trivial in the Markov operator and further increasing samples would result in negligible improvements. Ideally, one could inject known physics into the learning framework to impose constraints on the learned kernels. However, in our first attempt in extracting MZ operators, we only apply the physical symmetries in the learning and let the data to inform/reveal the structure of the MZ operators. Figure \ref{fig:converge_component} (b) shows similar results for the memory kernel $\bfvO_{\Delta}^{(1)}$, which provide evidence for the convergence of the learned memory kernel.

\subsection{Generalized Fluctuation-Dissipation Relation}
\label{sec:GFD}

\begin{figure}[hbt]
    \centering
    \includegraphics[width=6in]{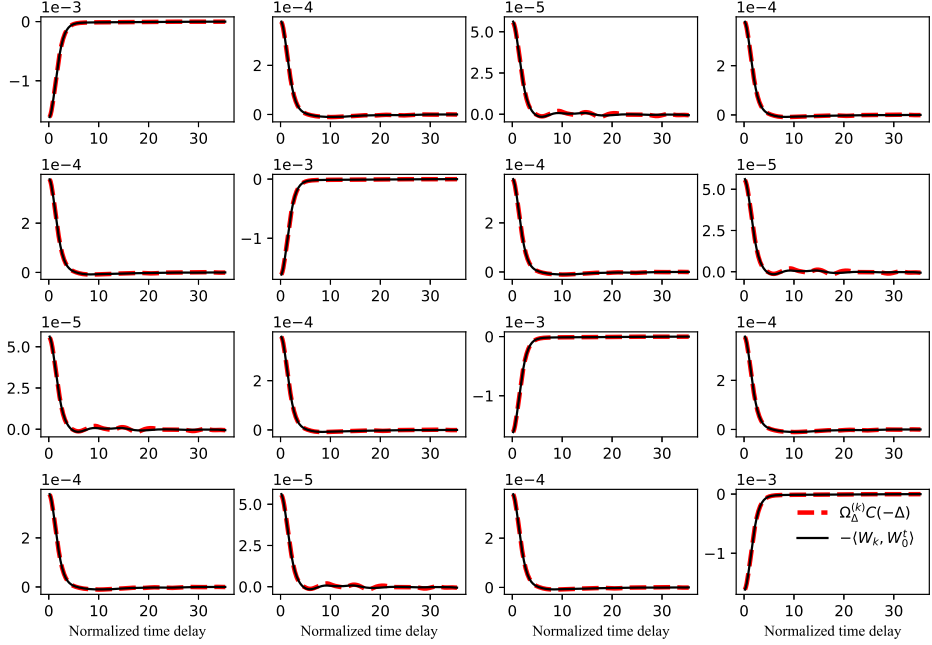}
    \caption{Numerical validation of the discrete-time GFD relation for $\Delta = 10dt$. The individual components of the matrices in the GFD relation (Eq.~\eqref{eqn:GFD}), $[\bfvO^{\l(l\r)}_\Delta \cdot \bfvC\l(-\Delta \r)]_{ij}$ and  $-[ \l\langle \bfvW_{l+1}, \bfvW_1^T\r\rangle]_{ij}$, $i,j \in 1,2,3,4$  are shown as a function of normalized time delay using Kolmogorv time scale. The LHS of Eq.~\eqref{eqn:GFD} is computed using the learned memory kernel , $\bfvO^{\l(l\r)}_\Delta$ and two-time correlation matrix $\bfvC(-\Delta)$. It is in very good agreement with the RHS: two-time correlation of the learned orthogonal dynamics $- \l\langle \bfvW_{l+1}, \bfvW_1^T\r\rangle$. }
    \label{fig:GFD}
\end{figure}

The Generalized Fluctuation-Dissipation relation (GFD) refers to the subtle self-consistent relationship between the learned memory kernel $\bfvO_\Delta^{\l(\ell\r)}$ and orthogonal dynamics $\bfvW_{l+1}$ with a suitable choice of projection operator. The specific GFD has been derived in \citet{lin2021data} for Mori's projection operator. This relation can be used to verify the correctness of the learned memory kernel and orthogonal dynamics. In figure \ref{fig:GFD}, we present the comparison between the left-hand-side (LHS) and the RHS of Eq.~\eqref{eqn:GFD}. The computed LHS aligns well with the RHS, confirming that the GFD relation is satisfied. 

\subsection{{Properties of the Learned Mori-Zwanzig Operators}}

There have been few attempts to perform quantitative analysis of the memory kernels and orthogonal dynamics of Navier-Stokes equations, due to the difficulty in developing tools for accurately and efficiently extracting them with nonlinear projection operators. The lack of knowledge makes it difficult to justify various assumptions on turbulence models based on MZ. In this section, we unveil some of the important properties of the extracted memory kernels and orthogonal dynamics with the current learning framework, in order to lay foundations for future turbulence model development.

\begin{figure}[hbt]
    \centering
    \includegraphics[width=6.5in]{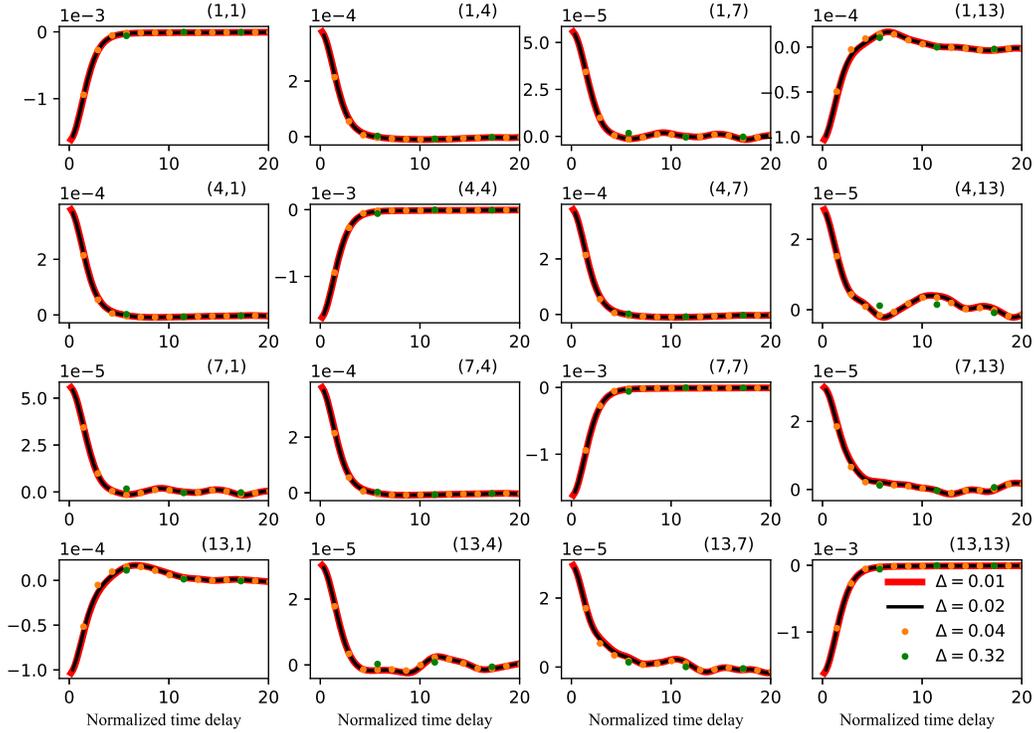}
    \caption{The components of the learned memory kernel for different discrete time intervals $\Delta=0.01, 0.02, 0.04, 0.032$.}
    \label{fig:fnorm_delta}
\end{figure}

In the discrete Mori-Zwanzig formulation, a hyper-parameter that is not related to the physical system is the discrete time interval $\Delta$. Given a database for a long trajectory with fixed temporal step $dt$, one can extract the corresponding Markov operator and memory kernels with the discrete time interval $\Delta$ being an integer number of the temporal step $\Delta=ndt$. \citet{lin2021data} showed that for the Lorenz '96 system, the memory kernels extracted with different $\Delta$ values collapse onto each other after applying a normalization factor $\Delta^{-2}$. Figure \ref{fig:fnorm_delta} shows the components of the extracted memory kernel (observable set 1) with different $\Delta$ values as a function of the time delay (normalized by Kolmogorov timescale $\tau_\eta$) for the N-S turbulence system. After proper normalization, the memory kernels extracted with different $\Delta$ values also collapse onto the same curve, despite a minute smoothing at the largest value ($\Delta=0.32$).
This shows that the structure of the memory kernel and memory length do not depend strongly on the discrete time interval $\Delta$.

\subsubsection{The Effects of Spatial Filters}
One of the most important assumptions in MZ-based models is about the length of the memory kernel. In the popular t-model \citep{hald2007optimal}, which has been applied to many nonlinear dynamical systems and achieved certain success, the convolution integral of the memory term is carried out using a simple left-hand quadrature rule, which can be interpreted as assuming an infinite long memory length. Other MZ-based models \citep{stinis2007higher,stinis2012mori,parish2017non,parish2017dynamic} also make various assumptions and approximations on the length and shape of the memory kernel.

\begin{figure}[hbt]
    \centering
    \includegraphics[width=6in]{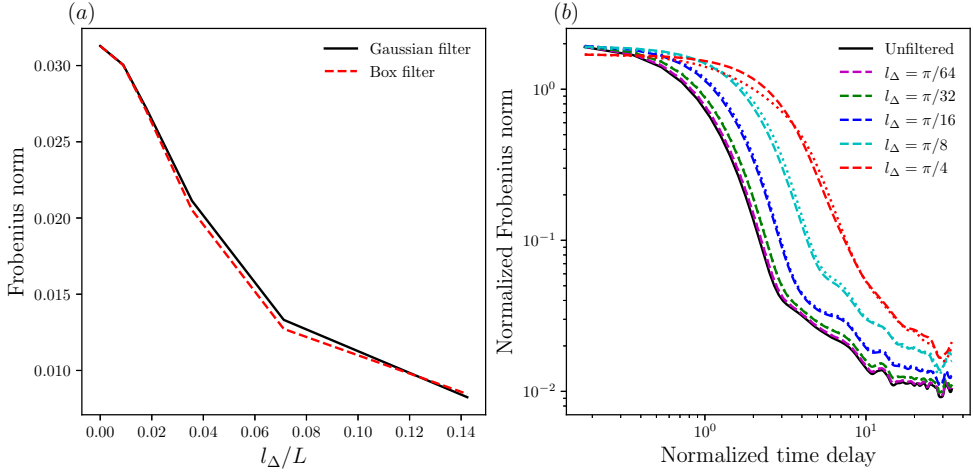}
    \caption{ (a) The Frobenius norm of Markov operator and (b) normalized Frobenius norm of the learned memory kernel for observable set 1, $\Vert\bfvO_\Delta^{\l(\ell\r)}\Vert_{F}$ as a function of normalized time delay. Two types of spatial filters, Gaussian and box filters, with various filtering length scales are applied to the physical-space variables. }
    \label{fig:fnorm_mem}
\end{figure}

In figure \ref{fig:fnorm_mem}, we present the Frobenius norm of Markov operator and the temporal decay of memory kernel. Two types of spatial filters that are commonly used in LES are also used here to compute the coarse-grained observables, in order to understand the effects of the filter type on memory length. In addition, the coarse-graining length scales as reflected by the filtering length $l_\Delta$ are also examined. Figure \ref{fig:fnorm_mem} (a) shows the dependence of Markovian contribution on the spatial filter sizes (normalized by integral length scale $L$). It is noted that the Markovian contribution decreases as the filtering size increases. In figure \ref{fig:fnorm_mem} (b), the Frobenius norm of memory kernel (normalized by its corresponding Markov operator) is plotted against the normalized time delay. From figure \ref{fig:fnorm_mem} (b), we can make a few important observations. First, the Frobenius norm of the memory kernel does not decrease to zero with a finite time delay; however, it becomes 2-3 orders of magnitude smaller at a time delay around several Kolmogorov timescales. This indicates that using finite support in the memory integral can be a reasonable modeling assumption because the contributions from large time delay are generally negligible. Second, the difference between the two different spatial filter types is small, but the effects of the filtering length scale are significant. With larger filtering sizes, the temporal decay of the memory kernel becomes slower, making the finite memory length longer, which indicates a shift of dynamical contributions from Markov term to memory integral.  Generally, when the filtering length scale increases, the range of scales that can be resolved by the coarse-grained observables becomes smaller. In this case, more past history of the observables needs to be included in the prediction, because the memory kernel formally characterizes the interactions between the coarse-grained dynamical variables and the under-resolved degrees of freedom.  These observations suggest a qualitative statement that the more we coarse-grain the observables, the less Markovian the coarse-grained model should be.

\begin{figure}[hbt]
    \centering
    \includegraphics[width=4in]{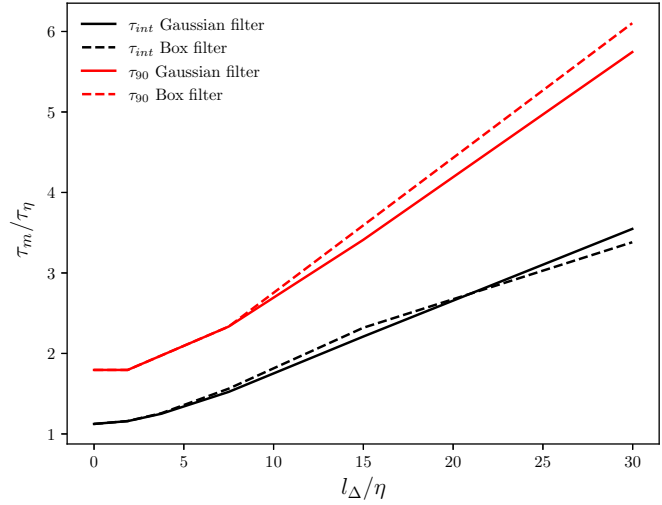}
    \caption{The dependence of timescale of the extracted memory kernel on the length scale of the filter. Two different methods are used to estimate the timescale: 1. $\tau_{int}$: timescale estimated using the integral of the memory kernel and 2. $\tau_{90}$ timescale estimated based on the location where the memory kernel drops to 10\% of the value at smallest time delay.}
    \label{fig:timescale_mem}
\end{figure}

So far, we have shown that the finite memory length assumption in the MZ-based models for N-S turbulence is generally reasonable. However, the quantitative estimation of such memory length/timescale can be challenging. Various approximation methods have been proposed to calculate the finite memory length to construct MZ-based models for turbulence. \citet{parish2017dynamic} used a dynamical procedure based on Germano identity \citep{germano1992turbulence} and \citet{parish2017non} used the spectral radius of the Jacobian of resolved variables to estimate the finite memory length. With the extracted memory kernels in the current study, we are able to calculate the timescales of the memory kernel and use them to verify the assumptions in these models. Figure \ref{fig:timescale_mem} shows the extracted timescales (normalized by Kolmogorov timescale) for various spatial filtering sizes (normalized by Kolmogorov scale). Two methods are employed for the calculation: 1) the integral of the memory kernel divided by the value at the smallest time delay, and 2) the time delay when the memory kernel dropped to 10\% of the maximum value. Note that we used a mean timescale based on the Frobenius norm to quantify the finite memory length, but the timescales may vary for different components of the memory kernel. We observe that the memory length is generally short, within the range of several Kolmogorov timescales. The memory length increases with the increase of the coarse-graining size, in agreement with those estimated by previously proposed models. There exists a weak dependence on the type of filters, which should be taken into account for modeling \citep{parish2017dynamic}.  We remark that it may seem that including memory integral may significantly increase the storage overhead, but when conducting reduced-order simulations using MZ-based models, it may not be necessary to store all the past information to calculate the memory kernel. Methods have been proposed based on the quadrature rule to model the memory integral as an additional set of ODEs \citep{stinis2012mori}, which are solved alongside the main coarse-grained equations. One of the aims of this work is to provide validation for constructing MZ-based memory closure models.

Addressing the noise/orthogonal dynamics is another essential component in developing models in Mori-Zwanzig formalism. Similar to the memory kernel, it is extremely difficult to theoretically solve the orthogonal dynamics due to its high-dimensionality and complexity, so the understanding of its property is still limited. To circumvent this difficulty, previous models based on MZ formalism have been focused on the projected image, where the noise term vanishes due to its orthogonality to the projection operator. Under the current learning framework, the noise term can be numerically extracted using the DNS database. By analyzing its property, we hope to shed some light on the modeling of the orthogonal dynamics. 

\begin{figure}[hbt]
    \centering
    \includegraphics[width=6in]{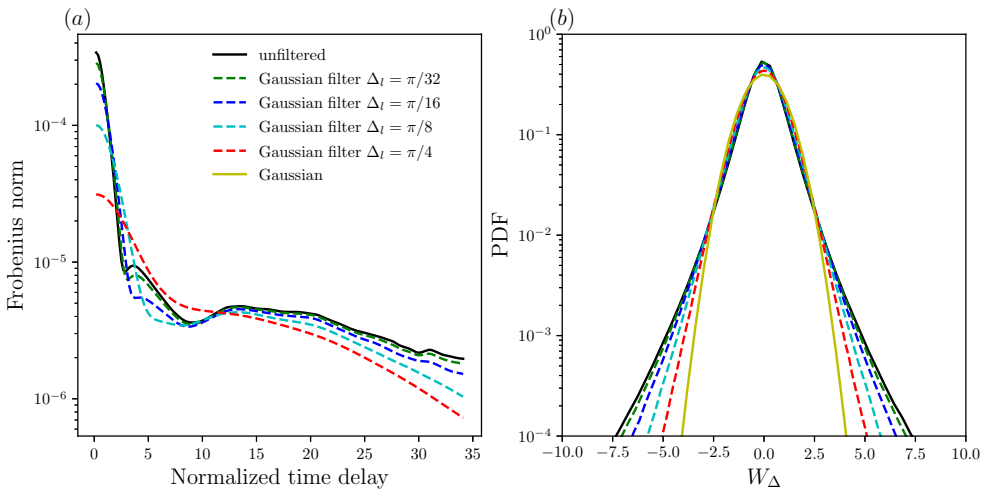}
    \caption{The Frobenious norm of the learned temporal correlation of the noise for observable set 1 (left), $\Vert\l\langle \bfvW_{l+1}, \bfvW_1^T\r\rangle\Vert_F$ as a function of time delay and PDFs of the orthogonal dynamics.}
    \label{fig:fnorm_noise}
\end{figure}

Figure \ref{fig:fnorm_noise} (a) shows the temporal correlation of the noise, as well as the PDFs of the orthogonal dynamics for the same set of observables as in figure \ref{fig:fnorm_mem}. We observe that there exists a similar temporal decay of the noise to the memory kernel. This behavior is expected, due to the existence of the GFD relation. Thus, Eq.~\eqref{eqn:GFD} states that the temporal correlation matrix of orthogonal dynamics $\l\langle \bfvW_{l+1}, \bfvW_1^T\r\rangle$ with a time delay $l\Delta$ should be equal to the memory kernel with same time delay multiply by a constant matrix $\bfvC(-\Delta)$, so it is reasonable for them to have qualitatively similar property. That is to say, if one seeks to derive a model for the noise with short temporal correlation, the corresponding memory kernel should have a similar length of memory kernel and structure. This property has been used to develop consistent MZ-based reduced-order models. On the other hand, the decay of the Frobenius norm of the two-time correlation of noise is not monotonic, different from that of the memory kernel. Between time delay of $5\tau_\eta \sim 15\tau_\eta$, there exists a transient behavior in the temporal decorrelation, which contains the nonlinear dynamics that can not be fully represented by the chosen observables in the linear space. This transient behavior becomes less significant with the increase of the filter sizes. If the correlation decays fast enough and the transient behavior happens after the fast decay and becomes trivial, the modeling can be greatly simplified by considering the correlation on a shorter timescale. Figure \ref{fig:fnorm_noise} (b) shows the normalized PDFs of the orthogonal dynamics with different filter sizes and a reference Gaussian distribution. The PDFs of the noise do not depend on the time delay (results not shown), so we only include in figure \ref{fig:fnorm_noise} the PDFs of the shortest time delay. With the increase of the filtering size, it can be seen that the noise becomes more Gaussian-like and there exists a trade-off between the timescale of the correlation and the "Gaussian-like" shape of the PDF.

\subsubsection{The Effects of Observables on the Extracted Operators}

\begin{figure}[hbt]
    \centering
    \includegraphics[width=6in]{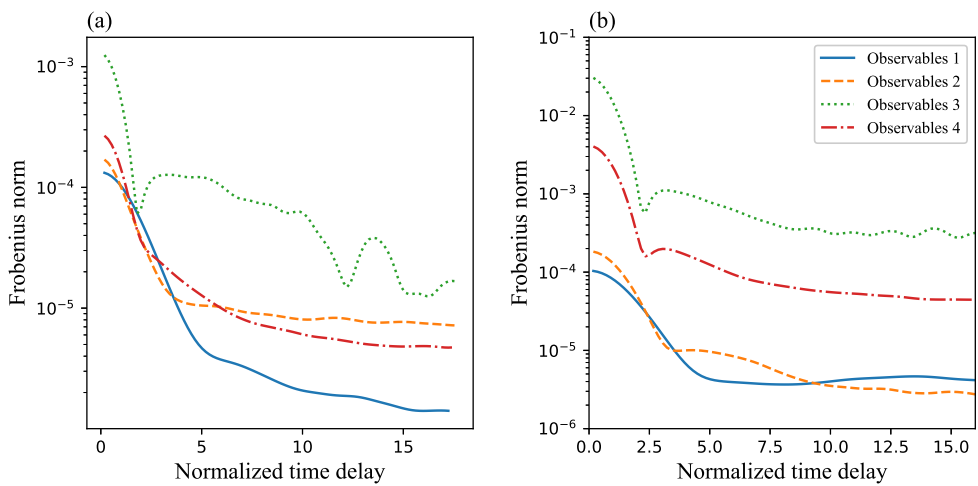}
    \caption{The Frobenius norm of the learned memory kernel (left) and two-time correlation of orthogonal dynamics (right) as a function of time delay.}
    \label{fig:fnorm_observables}
\end{figure}

Similar to Koopman learning, choosing the appropriate set of observables that can best close the original dynamics systems in the linear space is an important task in the current MZ learning framework, with another layer of complexity due to the existence of memory kernel and orthogonal dynamics. Here, we investigate the effects of different sets of observables on the properties of the learned kernels. These sets are described in \ref{sec:dnsdata}, and correspond to different aspects of N-S equations and turbulence variables. Figure \ref{fig:fnorm_observables} shows the Frobenius norm of the learned memory kernels for four sets of observables and the Frobenius norm of the corresponding orthogonal dynamics. It is obvious from figure \ref{fig:fnorm_observables} that the choices of observables for constructing MZ-based models can significantly influence the property of the extracted operators. The observable sets 2 and 4 have a similar magnitude of the memory kernel and the rate of decay as compared to observable set 1. On the other hand, observable set 3, which contains variables that are related to small-scale turbulence phenomena, exhibits a transient process of the memory effect: the faster early decay of the memory kernel is related to variables that are short-time correlated; the later increase then shows the intermittent behavior of the chosen observables. From the modeling perspective, it is more challenging to devise an accurate model that can reproduce the transient memory decay of the observable set 3. As for the property of the orthogonal dynamics, all four sets of observables show a certain level of transient behavior rather than exponential decay.

\begin{figure}[hbt]
    \centering
    \includegraphics[width=6in]{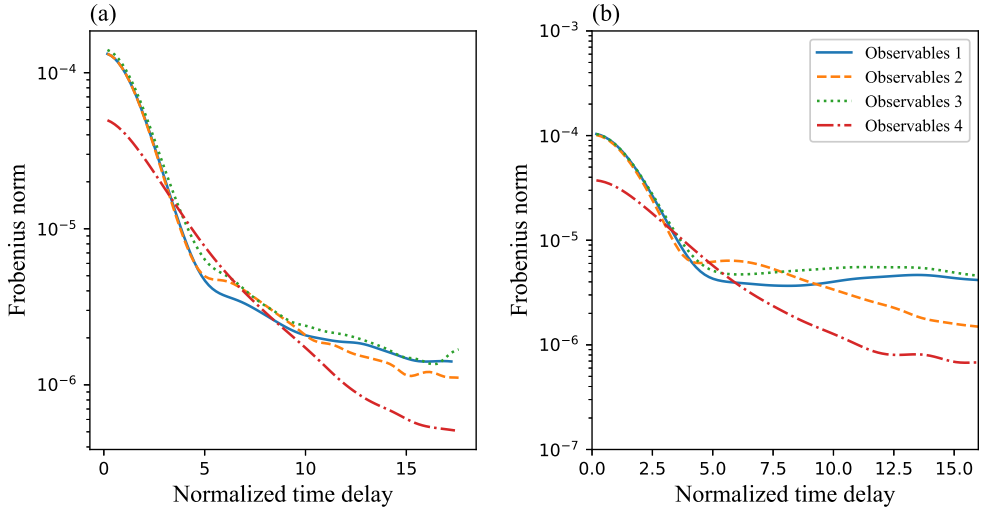}
    \caption{The Frobenius norm of the learned memory kernel (left) and two-time correlation of orthogonal dynamics (right) for the shared observables among the 4 sets (filtered velocity) as a function of time delay.}
    \label{fig:fnorm_observables_u}
\end{figure}

The memory kernels shown in figure \ref{fig:fnorm_observables} contain components from all the observables in each set, making it difficult to study how the quantities of interest, e.g., filtered velocities, are affected by the different choices of observables.  To understand this, we extract from the learned kernels the subset of the matrices that correspond to the filtered velocity and show them in figure \ref{fig:fnorm_observables_u}. We observe that the property of the memory kernel for the filtered velocity is not significantly modified by including more observables in sets 2 and 3, even though the memory kernels for the full set of observables exhibit significant differences. On the other hand, for the observable set 4, the contribution from memory kernel is reduced and the decay of the memory kernel becomes smoother. A similar trend is observed for the noise, with a smaller magnitude and smoother rate of decay. We conclude from these observations that by using observable set 4, which contains the RHS of the governing equations, the contribution to dynamics shifts from memory kernel to the Markov term. By comparing figure \ref{fig:fnorm_observables} and \ref{fig:fnorm_observables_u}, we note that different sets of observables may exhibit different memory kernel structure and memory length. This implies that the modeling strategy may need to be changed for different observables. We also remark that adding more observables may not be beneficial for improving the properties of memory kernel for modeling, as shown by observable set 3 results.

\subsection{Numerical Experiments on Memory Effects}

In this section, we conduct numerical experiments to illustrate the advantage of including memory kernels for prediction. Consider the following procedure: a) After we apply the learning algorithm for a set of observables $\bfvg$ with a time interval $\Delta$, we generate additional samples ($>2\times 10^4$) on the trajectories of the same observables using the fully resolved simulation. These samples had not been used in the learning of the corresponding kernels. Within each sample, the snapshots are also evenly spaced in time with the same time interval $\Delta$. b) The total length of the trajectories is longer than the time scale of the memory kernel. c) We then use the additional snapshots to predict $\delta=n\Delta$ into the future using the following formula recursively:

\begin{equation}
    \bfvg_{pred}((n+1)\Delta)  =   \bfvO_\Delta^{\l(0\r)} \cdot  \bfvg(n\Delta) + \sum_{\ell=1}^{n} \bfvO_\Delta^{\l(\ell\r)} \cdot  \bfvg((n-l)\Delta),
\end{equation}

\noindent, which is the discrete GLE (Eq.~\eqref{eqn:dmoriGLE}) with the assumption that the orthogonal dynamics $\bfvW_{n+1}=0$, and calculate the errors on the prediction. d) The errors are then averaged over samples to reduce statistical variability. In this work, we choose the $L^2$-norm as the measure of prediction error, which is calculated as:

\begin{equation}
    \epsilon^2=\Vert\bfvg_{pred}-\bfvg_{DNS}\Vert_2^2,
\end{equation}

\noindent where $\bfvg_{pred}$ and $\bfvg_{DNS}$ denote the observables from predictions and the ``ground truth'' DNS simulations.

\begin{figure}[hbt]
    \centering
    \includegraphics[width=7in]{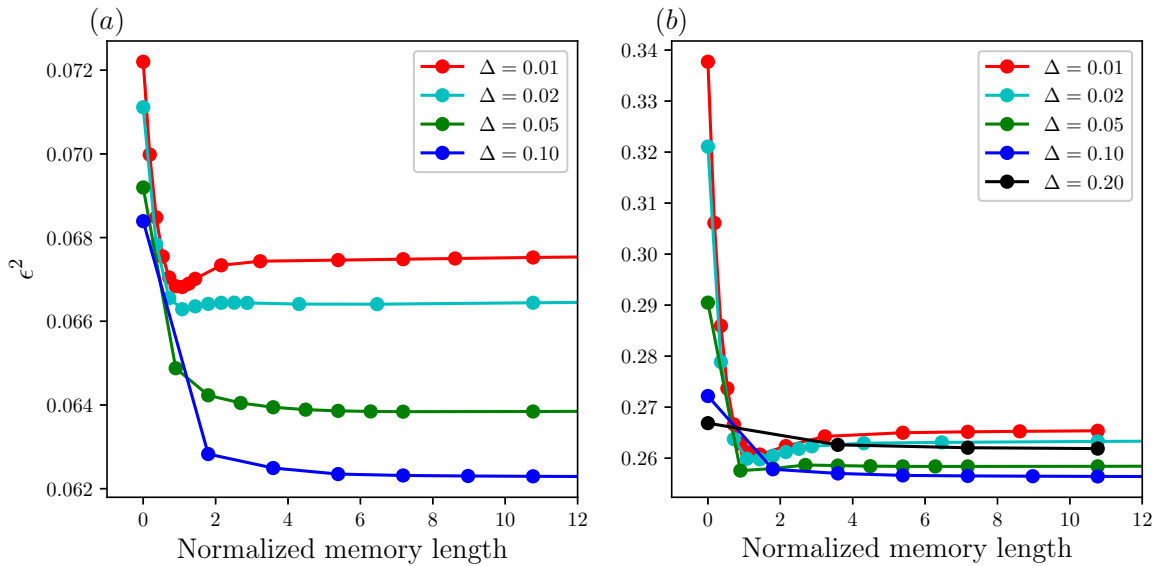}
    \caption{Comparison of prediction errors of different discrete time intervals $\Delta = 0.01, 0.02, 0.05, 0.10, 0.2$ as a function of the memory length. The prediction horizon is (a) $0.1$ and (b) $0.2$, which are multiples of the discrete time intervals. The error are calculated using the observables at the final prediction time.}
    \label{fig:error_delta}
\end{figure}

Figure \ref{fig:error_delta} shows the $L^2$-norm of the prediction errors using the discrete Mori-Zwanzig formulation with different memory lengths for prediction. The memory length is normalized using Kolmogorov timescale. We also consider different discrete time intervals $\Delta = 0.01, 0.02, 0.05, 0.1$ for learning the MZ operators. The chosen prediction horizons $\delta$ are $0.1$ and $0.2$, which are multiples of the discrete time intervals. We point out that the first point on the plot has memory length $0.0$ so that it corresponds to Markovian models and can be viewed as the Koopman prediction. It is evident from figure \ref{fig:error_delta} that the prediction errors decrease when past histories are included in the memory kernel.  As the memory length further increases past one Kolmogorov timescale, their effects on the prediction errors vary for different discrete time intervals $\Delta$: for the smallest time interval, the prediction error will increase and then saturate when the memory length is larger than $4\tau_\eta$. Our numerical results suggest the existence of an optimal memory length when the MZ operators are extracted with small time intervals $\Delta$. When the discrete time interval increases, the optimal memory length disappears and the prediction errors decrease and saturate after a certain memory length. The smallest prediction errors can be achieved when the discrete time interval $\Delta$ matches with the prediction horizon. For a longer prediction horizon $\delta=0.2$, similar observations can be made on the improvement of prediction errors by including past history. On the other hand, when the discrete time interval $\Delta$ matches with the prediction horizon, the error actually becomes larger than that of the smaller discrete time interval, which shows that there also exists an optimal discrete time interval for the selected prediction horizon. For the larger prediction horizon, the overall improvement by including past history is around 23\%. Note that the prediction errors depend on the accumulated magnitude of the orthogonal dynamics, which is neglected in the current prediction method. When the discrete time interval $\Delta$ is small, we need more steps to advance the observables so that the magnitude of the accumulated orthogonal dynamics that are missed in the prediction is larger. On the other hand, when the discrete time interval is too large (compared to the timescale of memory kernel), the projected image across such a large step can become small, which in turn will increase the magnitude of orthogonal dynamics. This may explain why the prediction error is large for discrete time interval $\Delta = 0.2$. Additionally, one should also take into account the different timescales of chosen observables, which may further complicate the process of choosing optimal discrete interval. Ideally, when proper models for orthogonal dynamics are included in prediction, one would not need to have this concern.

\begin{figure}[hbt]
    \centering
    \includegraphics[width=6in]{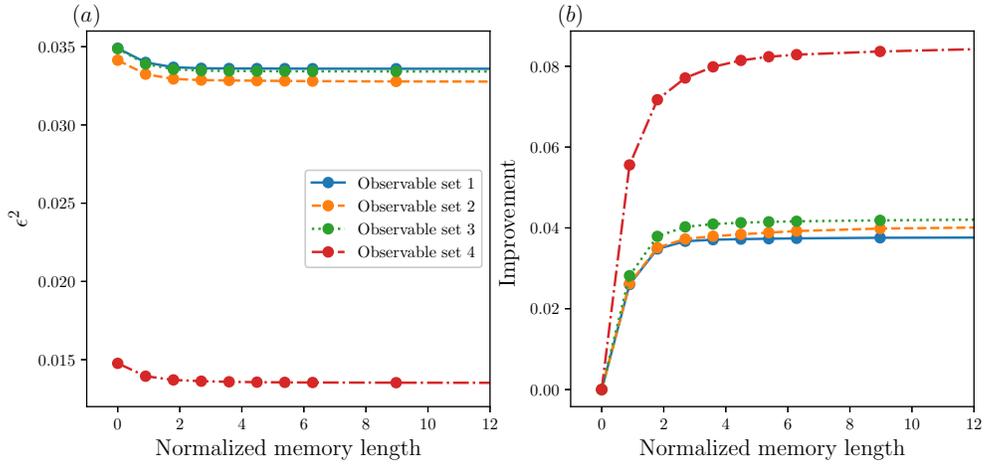}
    \caption{Comparison of prediction errors of different sets of observables as a function of the memory length. The prediction horizon is $0.05$, the same as the discrete time interval $\Delta$. The error are calculated using the shared observables (filtered velocity) at the final prediction time.}
    \label{fig:error_observable}
\end{figure}

The effects on prediction by choosing different sets of observables are explored next. Figure \ref{fig:error_observable} compares the prediction errors of the quantities of interest (filtered velocity) across the four sets of observables presented above. The prediction horizon is $\delta=0.05$ and it is the same as discrete time interval, so we only need to solve the discrete MZ formulation (without a model for orthogonal dynamics) for one step. In addition to the prediction improvement by including past history, the prediction errors vary drastically across the four sets of observables. Including turbulence physics-based observables like strain rate tensor, vorticity, and kinetic energy leads to negligible improvement on the prediction error for the observables of interest, namely filtered velocities. By including the sub-grid stress, a commonly used observable in traditional turbulence modeling approaches, we can observe a marginal improvement. The largest improvement can be seen for the observable set 4, where the observables are the terms in the filtered governing equations. There is over 50\% improvement on the prediction error by using observable set 4 compared to the other sets. The majority of the improvement by selecting observable set 4 comes from the Markov term, which is in good agreement with previous analyses of the extracted memory kernel. Further improvement in prediction can be achieved by including the past history in the memory term. The percentage improvement over the corresponding Markovian prediction method is also different for different sets of observables: around 4\% for observable set 1-3 and 8.5\% for the observable set 4. Overall, we can conclude that the choice of observables significantly affects the prediction capability of the learned MZ kernels.

Note that in the current numerical experiments, we neglect an important component of the MZ formalism, namely the orthogonal dynamics/noise. The orthogonal dynamics encode the important unresolved initial conditions of the dynamics that are important in reproducing the correct statistical property of the original system. In practice, it is more desirable to select a set of observables with a smaller magnitude of the orthogonal dynamics and shorter temporal correlation, which adds another layer of complexity in selecting observables for prediction. Our extracted noise exhibits complicated/nontrivial two-time correlation, meaning that the orthogonal dynamics can only be modeled by highly nontrivial color noise. 


\section{Conclusions}
\label{sec:conclusion}
Developing reduced-order models for turbulence is a challenging problem due to the existence of wide range scales. Traditional modeling strategies based on moment closure (RANS, LES) are derived based on physical intuition. On the other end of the spectrum, Mori-Zwanzig (MZ) formalism provides a formal mathematical procedure for derivation of low-dimensional representation of high-dimensional nonlinear dynamical systems. The outcome of applying the MZ formalism is the emergence of a memory term and orthogonal dynamics, which make solving the full low-dimensional system computationally expensive. Proper models for memory kernel and orthogonal dynamics need to be developed, which requires a comprehensive understanding of their mathematical properties. However, efforts of directly extracting the memory kernel and orthogonal dynamics in turbulent flows have been scarce, so the understanding of their properties is limited. In this work, we are the first to apply a data-driven algorithm to a fully-resolved turbulence simulation dataset to extract the memory kernel and orthogonal dynamics and analyze their properties. This provides a foundation for developing accurate MZ-based turbulence models including contributions from memory and orthogonal dynamics.

 With data augmentation using known turbulence symmetries, the Markov, memory kernels, and orthogonal dynamics can be successfully extracted using a reasonable amount of data and are shown to be statistically converged. The subtle Generalized Fluctuation Dissipation relation, which is a natural outcome of MZ formalism, is verified numerically using the extracted kernels and two-time correlation function of the orthogonal dynamics. This confirms the accuracy and correctness of the learning procedure and the learned MZ kernels. The properties of the memory kernels are shown to have a strong dependence on the spatial filtering sizes and weak dependence on filtering type. The Frobenius norm of the memory kernel exhibits a fast decay, indicating that the finite memory length modeling assumption is reasonable. The timescales of the memory kernels are then calculated and qualitative agreement can be observed with previous studies. The two-time correlation matrix of the orthogonal dynamics exhibits similar dependence on the spatial filtering size as the memory kernel. With a larger filter size, the PDF of orthogonal dynamics becomes more Gaussian-like. The effects of different observables on the extracted kernels are then examined. We observe that expanding the set of observables using nonlinear functions may significantly influence the decay of the memory kernel. On the other hand, the memory lengths of the added nonlinear observables are different, implying that a multi-timescale model may be needed for the expanded observable set. By using the observable set 4, which includes the right-hand-side of the filtered N-S equation, the magnitude of the memory kernel decreases. This is explained as a shift of contributions from memory term to Markov term.
 
 The advantages of including past history in prediction using MZ-based models are studied by comparing the prediction error with only the Markovian model. Results show that the $\bm{L}^2$ prediction error is lowered by including the memory integral, especially for longer predicting horizon. When using kernels extracted on a smaller discrete time interval, there exists an optimal memory length for calculating the memory integral. On the other hand, for larger discrete time interval, the improvement on prediction saturates when long enough past history is used. The optimal discrete time interval may also change for different prediction horizons, which is related to the relative magnitude of discrete time interval and memory length. Finally, the influence of observables on prediction is investigated. It is shown that the improvement on prediction of filtered velocity is marginal when including physics-based observables. By using equation-based observables (observable set 4), the improvement of the Markovian model is significant. In addition, including the memory integral further improves the prediction accuracy and the percentage increase in accuracy is also larger for observable set 4.
 In this study, we only show the improvements in prediction due to the inclusion of memory effects, but without the orthogonal dynamics. When proper models for orthogonal dynamics are proposed and used for prediction, the prediction should significantly improve and the concerns on choosing the optimal prediction parameters could be alleviated. Future works will be dedicated to developing MZ-based turbulence models from the following perspectives: (a) discovering observables with suitable properties for modeling (finite memory length and simple profile of the memory kernel) and (b) devising stochastic models for orthogonal dynamics that satisfies the learned statistical properties.

\begin{acknowledgments}
This work was performed under the auspices of DOE. Financial support comes from Los Alamos National Laboratory (LANL), Laboratory Directed Research and Development (LDRD) project "Machine Learning for Turbulence," 20190059DR. LANL, an affirmative action/equal opportunity employer, is managed by Triad National Security, LLC, for the National Nuclear Security Administration of the U.S. Department of Energy under contract 89233218CNA000001. YTL was partially supported by project LDRD-20190034ER (Massively Parallel Acceleration of the Dynamics of Complex Systems: a Data-Driven Approach.)
\end{acknowledgments}

\bibliography{main}

\begin{thebibliography}{34}%
\makeatletter
\providecommand \@ifxundefined [1]{%
 \@ifx{#1\undefined}
}%
\providecommand \@ifnum [1]{%
 \ifnum #1\expandafter \@firstoftwo
 \else \expandafter \@secondoftwo
 \fi
}%
\providecommand \@ifx [1]{%
 \ifx #1\expandafter \@firstoftwo
 \else \expandafter \@secondoftwo
 \fi
}%
\providecommand \natexlab [1]{#1}%
\providecommand \enquote  [1]{``#1''}%
\providecommand \bibnamefont  [1]{#1}%
\providecommand \bibfnamefont [1]{#1}%
\providecommand \citenamefont [1]{#1}%
\providecommand \href@noop [0]{\@secondoftwo}%
\providecommand \href [0]{\begingroup \@sanitize@url \@href}%
\providecommand \@href[1]{\@@startlink{#1}\@@href}%
\providecommand \@@href[1]{\endgroup#1\@@endlink}%
\providecommand \@sanitize@url [0]{\catcode `\\12\catcode `\$12\catcode
  `\&12\catcode `\#12\catcode `\^12\catcode `\_12\catcode `\%12\relax}%
\providecommand \@@startlink[1]{}%
\providecommand \@@endlink[0]{}%
\providecommand \url  [0]{\begingroup\@sanitize@url \@url }%
\providecommand \@url [1]{\endgroup\@href {#1}{\urlprefix }}%
\providecommand \urlprefix  [0]{URL }%
\providecommand \Eprint [0]{\href }%
\providecommand \doibase [0]{http://dx.doi.org/}%
\providecommand \selectlanguage [0]{\@gobble}%
\providecommand \bibinfo  [0]{\@secondoftwo}%
\providecommand \bibfield  [0]{\@secondoftwo}%
\providecommand \translation [1]{[#1]}%
\providecommand \BibitemOpen [0]{}%
\providecommand \bibitemStop [0]{}%
\providecommand \bibitemNoStop [0]{.\EOS\space}%
\providecommand \EOS [0]{\spacefactor3000\relax}%
\providecommand \BibitemShut  [1]{\csname bibitem#1\endcsname}%
\let\auto@bib@innerbib\@empty
\bibitem [{\citenamefont {Sagaut}(2006)}]{sagaut2006large}%
  \BibitemOpen
  \bibfield  {author} {\bibinfo {author} {\bibfnamefont {P.}~\bibnamefont
  {Sagaut}},\ }\href@noop {} {\emph {\bibinfo {title} {Large eddy simulation
  for incompressible flows: an introduction}}}\ (\bibinfo  {publisher}
  {Springer Science \& Business Media},\ \bibinfo {year} {2006})\BibitemShut
  {NoStop}%
\bibitem [{\citenamefont {Kraichnan}(1964)}]{kraichnan1964decay}%
  \BibitemOpen
  \bibfield  {author} {\bibinfo {author} {\bibfnamefont {R.~H.}\ \bibnamefont
  {Kraichnan}},\ }\bibfield  {title} {\enquote {\bibinfo {title} {Decay of
  isotropic turbulence in the direct-interaction approximation},}\ }\href@noop
  {} {\bibfield  {journal} {\bibinfo  {journal} {The Physics of Fluids}\
  }\textbf {\bibinfo {volume} {7}},\ \bibinfo {pages} {1030--1048} (\bibinfo
  {year} {1964})}\BibitemShut {NoStop}%
\bibitem [{\citenamefont {Kraichnan}(1965)}]{kraichnan1965lagrangian}%
  \BibitemOpen
  \bibfield  {author} {\bibinfo {author} {\bibfnamefont {R.~H.}\ \bibnamefont
  {Kraichnan}},\ }\bibfield  {title} {\enquote {\bibinfo {title}
  {Lagrangian-history closure approximation for turbulence},}\ }\href@noop {}
  {\bibfield  {journal} {\bibinfo  {journal} {The Physics of Fluids}\ }\textbf
  {\bibinfo {volume} {8}},\ \bibinfo {pages} {575--598} (\bibinfo {year}
  {1965})}\BibitemShut {NoStop}%
\bibitem [{\citenamefont {Bos}\ and\ \citenamefont
  {Bertoglio}(2013)}]{bos2013lagrangian}%
  \BibitemOpen
  \bibfield  {author} {\bibinfo {author} {\bibfnamefont {W.~J.}\ \bibnamefont
  {Bos}}\ and\ \bibinfo {author} {\bibfnamefont {J.-P.}\ \bibnamefont
  {Bertoglio}},\ }\bibfield  {title} {\enquote {\bibinfo {title} {Lagrangian
  markovianized field approximation for turbulence},}\ }\href@noop {}
  {\bibfield  {journal} {\bibinfo  {journal} {Journal of Turbulence}\ }\textbf
  {\bibinfo {volume} {14}},\ \bibinfo {pages} {99--120} (\bibinfo {year}
  {2013})}\BibitemShut {NoStop}%
\bibitem [{\citenamefont {Mori}(1965)}]{mori1965transport}%
  \BibitemOpen
  \bibfield  {author} {\bibinfo {author} {\bibfnamefont {H.}~\bibnamefont
  {Mori}},\ }\bibfield  {title} {\enquote {\bibinfo {title} {Transport,
  collective motion, and brownian motion},}\ }\href@noop {} {\bibfield
  {journal} {\bibinfo  {journal} {Progress of theoretical physics}\ }\textbf
  {\bibinfo {volume} {33}},\ \bibinfo {pages} {423--455} (\bibinfo {year}
  {1965})}\BibitemShut {NoStop}%
\bibitem [{\citenamefont {Zwanzig}(1973)}]{zwanzig1973nonlinear}%
  \BibitemOpen
  \bibfield  {author} {\bibinfo {author} {\bibfnamefont {R.}~\bibnamefont
  {Zwanzig}},\ }\bibfield  {title} {\enquote {\bibinfo {title} {Nonlinear
  generalized langevin equations},}\ }\href@noop {} {\bibfield  {journal}
  {\bibinfo  {journal} {Journal of Statistical Physics}\ }\textbf {\bibinfo
  {volume} {9}},\ \bibinfo {pages} {215--220} (\bibinfo {year}
  {1973})}\BibitemShut {NoStop}%
\bibitem [{\citenamefont {Chorin}, \citenamefont {Hald},\ and\ \citenamefont
  {Kupferman}(2000)}]{chorin2000optimal}%
  \BibitemOpen
  \bibfield  {author} {\bibinfo {author} {\bibfnamefont {A.~J.}\ \bibnamefont
  {Chorin}}, \bibinfo {author} {\bibfnamefont {O.~H.}\ \bibnamefont {Hald}}, \
  and\ \bibinfo {author} {\bibfnamefont {R.}~\bibnamefont {Kupferman}},\
  }\bibfield  {title} {\enquote {\bibinfo {title} {Optimal prediction and the
  mori--zwanzig representation of irreversible processes},}\ }\href@noop {}
  {\bibfield  {journal} {\bibinfo  {journal} {Proceedings of the National
  Academy of Sciences}\ }\textbf {\bibinfo {volume} {97}},\ \bibinfo {pages}
  {2968--2973} (\bibinfo {year} {2000})}\BibitemShut {NoStop}%
\bibitem [{\citenamefont {Givon}, \citenamefont {Kupferman},\ and\
  \citenamefont {Hald}(2005)}]{givon2005existence}%
  \BibitemOpen
  \bibfield  {author} {\bibinfo {author} {\bibfnamefont {D.}~\bibnamefont
  {Givon}}, \bibinfo {author} {\bibfnamefont {R.}~\bibnamefont {Kupferman}}, \
  and\ \bibinfo {author} {\bibfnamefont {O.~H.}\ \bibnamefont {Hald}},\
  }\bibfield  {title} {\enquote {\bibinfo {title} {Existence proof for
  orthogonal dynamics and the mori-zwanzig formalism},}\ }\href@noop {}
  {\bibfield  {journal} {\bibinfo  {journal} {Israel Journal of Mathematics}\
  }\textbf {\bibinfo {volume} {145}},\ \bibinfo {pages} {221--241} (\bibinfo
  {year} {2005})}\BibitemShut {NoStop}%
\bibitem [{\citenamefont {Chorin}\ and\ \citenamefont
  {Stinis}(2007)}]{chorin2007problem}%
  \BibitemOpen
  \bibfield  {author} {\bibinfo {author} {\bibfnamefont {A.}~\bibnamefont
  {Chorin}}\ and\ \bibinfo {author} {\bibfnamefont {P.}~\bibnamefont
  {Stinis}},\ }\bibfield  {title} {\enquote {\bibinfo {title} {Problem
  reduction, renormalization, and memory},}\ }\href@noop {} {\bibfield
  {journal} {\bibinfo  {journal} {Communications in Applied Mathematics and
  Computational Science}\ }\textbf {\bibinfo {volume} {1}},\ \bibinfo {pages}
  {1--27} (\bibinfo {year} {2007})}\BibitemShut {NoStop}%
\bibitem [{\citenamefont {Bernstein}(2007)}]{bernstein2007optimal}%
  \BibitemOpen
  \bibfield  {author} {\bibinfo {author} {\bibfnamefont {D.}~\bibnamefont
  {Bernstein}},\ }\bibfield  {title} {\enquote {\bibinfo {title} {Optimal
  prediction of burgers’s equation},}\ }\href@noop {} {\bibfield  {journal}
  {\bibinfo  {journal} {Multiscale Modeling \& Simulation}\ }\textbf {\bibinfo
  {volume} {6}},\ \bibinfo {pages} {27--52} (\bibinfo {year}
  {2007})}\BibitemShut {NoStop}%
\bibitem [{\citenamefont {Hald}\ and\ \citenamefont
  {Stinis}(2007)}]{hald2007optimal}%
  \BibitemOpen
  \bibfield  {author} {\bibinfo {author} {\bibfnamefont {O.~H.}\ \bibnamefont
  {Hald}}\ and\ \bibinfo {author} {\bibfnamefont {P.}~\bibnamefont {Stinis}},\
  }\bibfield  {title} {\enquote {\bibinfo {title} {Optimal prediction and the
  rate of decay for solutions of the euler equations in two and three
  dimensions},}\ }\href@noop {} {\bibfield  {journal} {\bibinfo  {journal}
  {Proceedings of the National Academy of Sciences}\ }\textbf {\bibinfo
  {volume} {104}},\ \bibinfo {pages} {6527--6532} (\bibinfo {year}
  {2007})}\BibitemShut {NoStop}%
\bibitem [{\citenamefont {Chandy}\ and\ \citenamefont
  {Frankel}(2010)}]{chandy2010t}%
  \BibitemOpen
  \bibfield  {author} {\bibinfo {author} {\bibfnamefont {A.~J.}\ \bibnamefont
  {Chandy}}\ and\ \bibinfo {author} {\bibfnamefont {S.~H.}\ \bibnamefont
  {Frankel}},\ }\bibfield  {title} {\enquote {\bibinfo {title} {The t-model as
  a large eddy simulation model for the navier--stokes equations},}\
  }\href@noop {} {\bibfield  {journal} {\bibinfo  {journal} {Multiscale
  Modeling \& Simulation}\ }\textbf {\bibinfo {volume} {8}},\ \bibinfo {pages}
  {445--462} (\bibinfo {year} {2010})}\BibitemShut {NoStop}%
\bibitem [{\citenamefont {Parish}\ and\ \citenamefont
  {Duraisamy}(2017{\natexlab{a}})}]{parish2017dynamic}%
  \BibitemOpen
  \bibfield  {author} {\bibinfo {author} {\bibfnamefont {E.~J.}\ \bibnamefont
  {Parish}}\ and\ \bibinfo {author} {\bibfnamefont {K.}~\bibnamefont
  {Duraisamy}},\ }\bibfield  {title} {\enquote {\bibinfo {title} {A dynamic
  subgrid scale model for large eddy simulations based on the mori--zwanzig
  formalism},}\ }\href@noop {} {\bibfield  {journal} {\bibinfo  {journal} {J.
  Comput. Phys.}\ }\textbf {\bibinfo {volume} {349}},\ \bibinfo {pages}
  {154--175} (\bibinfo {year} {2017}{\natexlab{a}})}\BibitemShut {NoStop}%
\bibitem [{\citenamefont {Stinis}(2013)}]{stinis2013renormalized}%
  \BibitemOpen
  \bibfield  {author} {\bibinfo {author} {\bibfnamefont {P.}~\bibnamefont
  {Stinis}},\ }\bibfield  {title} {\enquote {\bibinfo {title} {Renormalized
  reduced models for singular pdes},}\ }\href@noop {} {\bibfield  {journal}
  {\bibinfo  {journal} {Communications in Applied Mathematics and Computational
  Science}\ }\textbf {\bibinfo {volume} {8}},\ \bibinfo {pages} {39--66}
  (\bibinfo {year} {2013})}\BibitemShut {NoStop}%
\bibitem [{\citenamefont {Stinis}(2015)}]{stinis2015renormalized}%
  \BibitemOpen
  \bibfield  {author} {\bibinfo {author} {\bibfnamefont {P.}~\bibnamefont
  {Stinis}},\ }\bibfield  {title} {\enquote {\bibinfo {title} {Renormalized
  mori--zwanzig-reduced models for systems without scale separation},}\
  }\href@noop {} {\bibfield  {journal} {\bibinfo  {journal} {Proceedings of the
  Royal Society A: Mathematical, Physical and Engineering Sciences}\ }\textbf
  {\bibinfo {volume} {471}},\ \bibinfo {pages} {20140446} (\bibinfo {year}
  {2015})}\BibitemShut {NoStop}%
\bibitem [{\citenamefont {Stinis}(2007)}]{stinis2007higher}%
  \BibitemOpen
  \bibfield  {author} {\bibinfo {author} {\bibfnamefont {P.}~\bibnamefont
  {Stinis}},\ }\bibfield  {title} {\enquote {\bibinfo {title} {Higher order
  mori--zwanzig models for the euler equations},}\ }\href@noop {} {\bibfield
  {journal} {\bibinfo  {journal} {Multiscale Modeling \& Simulation}\ }\textbf
  {\bibinfo {volume} {6}},\ \bibinfo {pages} {741--760} (\bibinfo {year}
  {2007})}\BibitemShut {NoStop}%
\bibitem [{\citenamefont {Parish}\ and\ \citenamefont
  {Duraisamy}(2017{\natexlab{b}})}]{parish2017non}%
  \BibitemOpen
  \bibfield  {author} {\bibinfo {author} {\bibfnamefont {E.~J.}\ \bibnamefont
  {Parish}}\ and\ \bibinfo {author} {\bibfnamefont {K.}~\bibnamefont
  {Duraisamy}},\ }\bibfield  {title} {\enquote {\bibinfo {title} {Non-markovian
  closure models for large eddy simulations using the mori-zwanzig
  formalism},}\ }\href@noop {} {\bibfield  {journal} {\bibinfo  {journal}
  {Phys. Rev. Fluids}\ }\textbf {\bibinfo {volume} {2}},\ \bibinfo {pages}
  {014604} (\bibinfo {year} {2017}{\natexlab{b}})}\BibitemShut {NoStop}%
\bibitem [{\citenamefont {Gouasmi}, \citenamefont {Parish},\ and\ \citenamefont
  {Duraisamy}(2017)}]{gouasmi2017priori}%
  \BibitemOpen
  \bibfield  {author} {\bibinfo {author} {\bibfnamefont {A.}~\bibnamefont
  {Gouasmi}}, \bibinfo {author} {\bibfnamefont {E.~J.}\ \bibnamefont {Parish}},
  \ and\ \bibinfo {author} {\bibfnamefont {K.}~\bibnamefont {Duraisamy}},\
  }\bibfield  {title} {\enquote {\bibinfo {title} {A priori estimation of
  memory effects in reduced-order models of nonlinear systems using the
  mori--zwanzig formalism},}\ }\href@noop {} {\bibfield  {journal} {\bibinfo
  {journal} {Proceedings of the Royal Society A: Mathematical, Physical and
  Engineering Sciences}\ }\textbf {\bibinfo {volume} {473}},\ \bibinfo {pages}
  {20170385} (\bibinfo {year} {2017})}\BibitemShut {NoStop}%
\bibitem [{\citenamefont {Koopman}(1931)}]{koopman1931hamiltonian}%
  \BibitemOpen
  \bibfield  {author} {\bibinfo {author} {\bibfnamefont {B.~O.}\ \bibnamefont
  {Koopman}},\ }\bibfield  {title} {\enquote {\bibinfo {title} {Hamiltonian
  systems and transformation in hilbert space},}\ }\href@noop {} {\bibfield
  {journal} {\bibinfo  {journal} {Proceedings of the national academy of
  sciences of the united states of america}\ }\textbf {\bibinfo {volume}
  {17}},\ \bibinfo {pages} {315} (\bibinfo {year} {1931})}\BibitemShut
  {NoStop}%
\bibitem [{\citenamefont {Koopman}\ and\ \citenamefont
  {Neumann}(1932)}]{koopman1932dynamical}%
  \BibitemOpen
  \bibfield  {author} {\bibinfo {author} {\bibfnamefont {B.}~\bibnamefont
  {Koopman}}\ and\ \bibinfo {author} {\bibfnamefont {J.~v.}\ \bibnamefont
  {Neumann}},\ }\bibfield  {title} {\enquote {\bibinfo {title} {Dynamical
  systems of continuous spectra},}\ }\href@noop {} {\bibfield  {journal}
  {\bibinfo  {journal} {Proceedings of the National Academy of Sciences of the
  United States of America}\ }\textbf {\bibinfo {volume} {18}},\ \bibinfo
  {pages} {255} (\bibinfo {year} {1932})}\BibitemShut {NoStop}%
\bibitem [{\citenamefont {Schmid}(2010)}]{schmid2010dynamic}%
  \BibitemOpen
  \bibfield  {author} {\bibinfo {author} {\bibfnamefont {P.~J.}\ \bibnamefont
  {Schmid}},\ }\bibfield  {title} {\enquote {\bibinfo {title} {Dynamic mode
  decomposition of numerical and experimental data},}\ }\href@noop {}
  {\bibfield  {journal} {\bibinfo  {journal} {Journal of fluid mechanics}\
  }\textbf {\bibinfo {volume} {656}},\ \bibinfo {pages} {5--28} (\bibinfo
  {year} {2010})}\BibitemShut {NoStop}%
\bibitem [{\citenamefont {Williams}, \citenamefont {Kevrekidis},\ and\
  \citenamefont {Rowley}(2015)}]{williams2015data}%
  \BibitemOpen
  \bibfield  {author} {\bibinfo {author} {\bibfnamefont {M.~O.}\ \bibnamefont
  {Williams}}, \bibinfo {author} {\bibfnamefont {I.~G.}\ \bibnamefont
  {Kevrekidis}}, \ and\ \bibinfo {author} {\bibfnamefont {C.~W.}\ \bibnamefont
  {Rowley}},\ }\bibfield  {title} {\enquote {\bibinfo {title} {A data--driven
  approximation of the koopman operator: Extending dynamic mode
  decomposition},}\ }\href@noop {} {\bibfield  {journal} {\bibinfo  {journal}
  {Journal of Nonlinear Science}\ }\textbf {\bibinfo {volume} {25}},\ \bibinfo
  {pages} {1307--1346} (\bibinfo {year} {2015})}\BibitemShut {NoStop}%
\bibitem [{\citenamefont {Lin}\ \emph {et~al.}(2021)\citenamefont {Lin},
  \citenamefont {Tian}, \citenamefont {Livescu},\ and\ \citenamefont
  {Anghel}}]{lin2021data}%
  \BibitemOpen
  \bibfield  {author} {\bibinfo {author} {\bibfnamefont {Y.~T.}\ \bibnamefont
  {Lin}}, \bibinfo {author} {\bibfnamefont {Y.}~\bibnamefont {Tian}}, \bibinfo
  {author} {\bibfnamefont {D.}~\bibnamefont {Livescu}}, \ and\ \bibinfo
  {author} {\bibfnamefont {M.}~\bibnamefont {Anghel}},\ }\bibfield  {title}
  {\enquote {\bibinfo {title} {Data-driven learning for the mori--zwanzig
  formalism: a generalization of the koopman learning framework},}\ }\href@noop
  {} {\bibfield  {journal} {\bibinfo  {journal} {arXiv preprint
  arXiv:2101.05873}\ } (\bibinfo {year} {2021})}\BibitemShut {NoStop}%
\bibitem [{\citenamefont {J~Evans}\ and\ \citenamefont
  {P~Morriss}(2007)}]{j2007statistical}%
  \BibitemOpen
  \bibfield  {author} {\bibinfo {author} {\bibfnamefont {D.}~\bibnamefont
  {J~Evans}}\ and\ \bibinfo {author} {\bibfnamefont {G.}~\bibnamefont
  {P~Morriss}},\ }\href@noop {} {\emph {\bibinfo {title} {Statistical mechanics
  of nonequilbrium liquids}}}\ (\bibinfo  {publisher} {ANU Press},\ \bibinfo
  {year} {2007})\BibitemShut {NoStop}%
\bibitem [{\citenamefont {Chorin}, \citenamefont {Hald},\ and\ \citenamefont
  {Kupferman}(2002)}]{chorin2002optimal}%
  \BibitemOpen
  \bibfield  {author} {\bibinfo {author} {\bibfnamefont {A.~J.}\ \bibnamefont
  {Chorin}}, \bibinfo {author} {\bibfnamefont {O.~H.}\ \bibnamefont {Hald}}, \
  and\ \bibinfo {author} {\bibfnamefont {R.}~\bibnamefont {Kupferman}},\
  }\bibfield  {title} {\enquote {\bibinfo {title} {Optimal prediction with
  memory},}\ }\href@noop {} {\bibfield  {journal} {\bibinfo  {journal} {Physica
  D: Nonlinear Phenomena}\ }\textbf {\bibinfo {volume} {166}},\ \bibinfo
  {pages} {239--257} (\bibinfo {year} {2002})}\BibitemShut {NoStop}%
\bibitem [{\citenamefont {Falkena}\ \emph {et~al.}(2019)\citenamefont
  {Falkena}, \citenamefont {Quinn}, \citenamefont {Sieber}, \citenamefont
  {Frank},\ and\ \citenamefont {Dijkstra}}]{falkena2019derivation}%
  \BibitemOpen
  \bibfield  {author} {\bibinfo {author} {\bibfnamefont {S.~K.}\ \bibnamefont
  {Falkena}}, \bibinfo {author} {\bibfnamefont {C.}~\bibnamefont {Quinn}},
  \bibinfo {author} {\bibfnamefont {J.}~\bibnamefont {Sieber}}, \bibinfo
  {author} {\bibfnamefont {J.}~\bibnamefont {Frank}}, \ and\ \bibinfo {author}
  {\bibfnamefont {H.~A.}\ \bibnamefont {Dijkstra}},\ }\bibfield  {title}
  {\enquote {\bibinfo {title} {Derivation of delay equation climate models
  using the mori-zwanzig formalism},}\ }\href@noop {} {\bibfield  {journal}
  {\bibinfo  {journal} {Proceedings of the Royal Society A}\ }\textbf {\bibinfo
  {volume} {475}},\ \bibinfo {pages} {20190075} (\bibinfo {year}
  {2019})}\BibitemShut {NoStop}%
\bibitem [{\citenamefont {Lin}\ and\ \citenamefont {Lu}(2019)}]{lin2019data}%
  \BibitemOpen
  \bibfield  {author} {\bibinfo {author} {\bibfnamefont {K.~K.}\ \bibnamefont
  {Lin}}\ and\ \bibinfo {author} {\bibfnamefont {F.}~\bibnamefont {Lu}},\
  }\bibfield  {title} {\enquote {\bibinfo {title} {Data-driven model reduction,
  wiener projections, and the mori-zwanzig formalism},}\ }\href@noop {}
  {\bibfield  {journal} {\bibinfo  {journal} {arXiv preprint arXiv:1908.07725}\
  } (\bibinfo {year} {2019})}\BibitemShut {NoStop}%
\bibitem [{\citenamefont {Smagorinsky}(1963)}]{smagorinsky1963general}%
  \BibitemOpen
  \bibfield  {author} {\bibinfo {author} {\bibfnamefont {J.}~\bibnamefont
  {Smagorinsky}},\ }\bibfield  {title} {\enquote {\bibinfo {title} {General
  circulation experiments with the primitive equations: I. the basic
  experiment},}\ }\href@noop {} {\bibfield  {journal} {\bibinfo  {journal}
  {Monthly weather review}\ }\textbf {\bibinfo {volume} {91}},\ \bibinfo
  {pages} {99--164} (\bibinfo {year} {1963})}\BibitemShut {NoStop}%
\bibitem [{\citenamefont {Yoshizawa}\ and\ \citenamefont
  {Horiuti}(1985)}]{yoshizawa1985statistically}%
  \BibitemOpen
  \bibfield  {author} {\bibinfo {author} {\bibfnamefont {A.}~\bibnamefont
  {Yoshizawa}}\ and\ \bibinfo {author} {\bibfnamefont {K.}~\bibnamefont
  {Horiuti}},\ }\bibfield  {title} {\enquote {\bibinfo {title} {A
  statistically-derived subgrid-scale kinetic energy model for the large-eddy
  simulation of turbulent flows},}\ }\href@noop {} {\bibfield  {journal}
  {\bibinfo  {journal} {Journal of the Physical Society of Japan}\ }\textbf
  {\bibinfo {volume} {54}},\ \bibinfo {pages} {2834--2839} (\bibinfo {year}
  {1985})}\BibitemShut {NoStop}%
\bibitem [{\citenamefont {Germano}\ \emph {et~al.}(1991)\citenamefont
  {Germano}, \citenamefont {Piomelli}, \citenamefont {Moin},\ and\
  \citenamefont {Cabot}}]{germano1991dynamic}%
  \BibitemOpen
  \bibfield  {author} {\bibinfo {author} {\bibfnamefont {M.}~\bibnamefont
  {Germano}}, \bibinfo {author} {\bibfnamefont {U.}~\bibnamefont {Piomelli}},
  \bibinfo {author} {\bibfnamefont {P.}~\bibnamefont {Moin}}, \ and\ \bibinfo
  {author} {\bibfnamefont {W.~H.}\ \bibnamefont {Cabot}},\ }\bibfield  {title}
  {\enquote {\bibinfo {title} {A dynamic subgrid-scale eddy viscosity model},}\
  }\href@noop {} {\bibfield  {journal} {\bibinfo  {journal} {Physics of Fluids
  A: Fluid Dynamics}\ }\textbf {\bibinfo {volume} {3}},\ \bibinfo {pages}
  {1760--1765} (\bibinfo {year} {1991})}\BibitemShut {NoStop}%
\bibitem [{\citenamefont {Petersen}\ and\ \citenamefont
  {Livescu}(2010)}]{petersen2010forcing}%
  \BibitemOpen
  \bibfield  {author} {\bibinfo {author} {\bibfnamefont {M.~R.}\ \bibnamefont
  {Petersen}}\ and\ \bibinfo {author} {\bibfnamefont {D.}~\bibnamefont
  {Livescu}},\ }\bibfield  {title} {\enquote {\bibinfo {title} {Forcing for
  statistically stationary compressible isotropic turbulence},}\ }\href@noop {}
  {\bibfield  {journal} {\bibinfo  {journal} {Physics of Fluids}\ }\textbf
  {\bibinfo {volume} {22}},\ \bibinfo {pages} {116101} (\bibinfo {year}
  {2010})}\BibitemShut {NoStop}%
\bibitem [{\citenamefont {Daniel}, \citenamefont {Livescu},\ and\ \citenamefont
  {Ryu}(2018)}]{daniel2018reaction}%
  \BibitemOpen
  \bibfield  {author} {\bibinfo {author} {\bibfnamefont {D.}~\bibnamefont
  {Daniel}}, \bibinfo {author} {\bibfnamefont {D.}~\bibnamefont {Livescu}}, \
  and\ \bibinfo {author} {\bibfnamefont {J.}~\bibnamefont {Ryu}},\ }\bibfield
  {title} {\enquote {\bibinfo {title} {Reaction analogy based forcing for
  incompressible scalar turbulence},}\ }\href@noop {} {\bibfield  {journal}
  {\bibinfo  {journal} {Physical Review Fluids}\ }\textbf {\bibinfo {volume}
  {3}},\ \bibinfo {pages} {094602} (\bibinfo {year} {2018})}\BibitemShut
  {NoStop}%
\bibitem [{\citenamefont {Stinis}(2012)}]{stinis2012mori}%
  \BibitemOpen
  \bibfield  {author} {\bibinfo {author} {\bibfnamefont {P.}~\bibnamefont
  {Stinis}},\ }\bibfield  {title} {\enquote {\bibinfo {title} {Mori-zwanzig
  reduced models for uncertainty quantification i: Parametric uncertainty},}\
  }\href@noop {} {\bibfield  {journal} {\bibinfo  {journal} {arXiv preprint
  arXiv:1211.4285}\ } (\bibinfo {year} {2012})}\BibitemShut {NoStop}%
\bibitem [{\citenamefont {Germano}\ \emph {et~al.}(1992)\citenamefont {Germano}
  \emph {et~al.}}]{germano1992turbulence}%
  \BibitemOpen
  \bibfield  {author} {\bibinfo {author} {\bibfnamefont {M.}~\bibnamefont
  {Germano}} \emph {et~al.},\ }\bibfield  {title} {\enquote {\bibinfo {title}
  {Turbulence- the filtering approach},}\ }\href@noop {} {\bibfield  {journal}
  {\bibinfo  {journal} {Journal of Fluid Mechanics}\ }\textbf {\bibinfo
  {volume} {238}},\ \bibinfo {pages} {325--336} (\bibinfo {year}
  {1992})}\BibitemShut {NoStop}%
\end{thebibliography}%
\end{document}